\documentclass[a4paper,11pt]{article}
\usepackage[utf8]{inputenc}
\usepackage[T1]{fontenc}
\usepackage{lmodern}
\usepackage[a4paper]{geometry}
\usepackage{color}
\usepackage{amsmath}
\usepackage{amssymb}
\usepackage{graphicx}

\usepackage{esint}
\usepackage[square,numbers,comma,sort&compress]{natbib}
\usepackage[unicode=true,bookmarks=false,breaklinks=false,pdfborder={0 0 1},backref=section,colorlinks=true] {hyperref}
\hypersetup{
 citecolor=blue,linkcolor=purple,urlcolor=purple}
\usepackage{breakurl}

\makeatletter

\providecommand{\tabularnewline}{\\}

\pdfoutput=1
\usepackage[usenames,dvipsnames,svgnames,table]{xcolor}
\usepackage{slashed}

\def\varabstract{ }
\def\varkeywords{ }
\def\vararxivnumber{ }
\def\vartitle{ }
\def\varsubtitle{ }
\renewcommand{\title}[1]{\gdef\vartitle{#1}}

\renewcommand{\abstract}[1]{\gdef\varabstract{#1}}
\newcommand{\keywords}[1]{\gdef\varkeywords{#1}}
\newcommand{\arxivnumber}[1]{\gdef\vararxivnumber{#1}}
\newtoks\authtoks
\renewcommand{\author}[2][]{%
\authtoks=\expandafter{\the\authtoks#2$^{#1}$\ }%
}
\newtoks\affiltoks
\newcommand{\affiliation}[2][]{%
    \affiltoks=\expandafter{\the\affiltoks{\item[$^{#1}$]#2}}%
}
\newtoks\emailtoks\newcounter{emailcounter}%
\newcommand{\emailAdd}[1]{%
\ifnum\theemailcounter>0\emailtoks=\expandafter{\the\emailtoks, \typeemail{#1}}%
\else\emailtoks=\expandafter{\typeemail{#1}}%
\fi
\stepcounter{emailcounter}}
\newcommand{\typeemail}[1]{\href{mailto:#1}{\tt #1}}

\renewcommand{\maketitle}{
\newgeometry{margin=2cm}
\pagestyle{empty}\setcounter{page}{0}
{\huge\flushleft\sffamily\bfseries\vartitle\\\Large\varsubtitle\par}
\vskip6ex
{\large\bfseries\raggedright\sffamily\the\authtoks\par}
\vskip2ex
\begin{list}{}{%
\setlength{\leftmargin}{0.28cm}%
\setlength{\labelsep}{0pt}%
\setlength{\itemsep}{-3pt}%
\setlength{\topsep}{-\parskip}}
\itshape\small%
\the\affiltoks
\end{list}
\vskip2ex
\noindent\hspace{0.28cm}\begin{minipage}[l]{.9\textwidth}
\begin{flushleft}
\textit{E-mail:} \the\emailtoks
\end{flushleft}
\end{minipage}
\vskip5ex
\noindent{\renewcommand\baselinestretch{.9}\textsc{Abstract:}}\ \varabstract
\vskip5ex 
\if!\varkeywords!\else\noindent{\textsc{Keywords:}}\ \varkeywords \vskip2ex\fi
\if!\vararxivnumber!\else\noindent{\textsc{ArXiv ePrint:}} \href{http://arxiv.org/abs/\vararxivnumber}{\vararxivnumber}\vskip2ex\fi

\newpage
\restoregeometry
\pagestyle{plain}

\setcounter{footnote}{0}
} 

\definecolor{MS}{rgb}{0,0,1}

\usepackage{ifthen}

\newcommand{\barlimc}[7]{
  \pgfmathparse{\mypos+0.3}
  \edef\mypos{\pgfmathresult}
\node[left,scale=0.6] at (0,\mypos) {#1};
\pgfmathparse{#3 > 5 ? 1 : 0}
\ifthenelse{\pgfmathresult=1}{
\fill[#2] ($(0,\mypos)+(0,-0.1)$) rectangle +(5,0.2);
\fill[white] ($(0,\mypos)+(3.5,-0.1)$) rectangle +(0.3,0.2);
\draw[decoration={zigzag},decorate,#2,very thick] (3.4,\mypos) to +(0.5,0);
\node[left,scale=0.6] at (5,\mypos) {#3};
}{
\fill[#2] ($(0,\mypos)+(0,-0.1)$) rectangle +(#3,0.2);
\node[left,scale=0.6] at (#3,\mypos) {#3};
}		
\fill[#4] ($(0,\mypos)+(0,-0.1)$) rectangle +(#5,0.2);
\node[left,scale=0.6] at (#5,\mypos) {#5};
\fill[#6] ($(0,\mypos)+(0,-0.1)$) rectangle +(#7,0.2);
\pgfmathparse{#7 <0.3 ? 1 : 0}
\ifthenelse{\pgfmathresult=1}{
\node[right,scale=0.6] at (0,\mypos) {#7};
}{
\node[left,scale=0.6] at (#7,\mypos) {#7};
}
}

\title{Beyond $\mathcal{R}(D^{(*)})$ with the general 2HDM-III for $b\to c\tau\nu$}

\author[1]{R. Martinez,}\emailAdd{remartinezm@unal.edu.co}
\author[2]{C.F. Sierra, }\emailAdd{cristian.sierra@monash.edu}
\author[2]{and German Valencia}\emailAdd{german.valencia@monash.edu}

\affiliation[1]{\textit{Departamento de F\'isica, Universidad Nacional de Colombia,
Ciudad Universitaria, K. 45 No. 26-85, Bogot\'a D.C., Colombia} }

\affiliation[2]{\textit{School of Physics and Astronomy, Monash University,
Melbourne VIC-3800}}

\abstract{
We review the parameter regions allowed by measurements of $\mathcal{R}(D^{(*)})$ 
and by a theoretical limit on ${\cal B}(B_{c}\to\tau\nu)$ in terms of   
generic scalar and pseudoscalar new physics couplings, $g_s$ and $g_p$. We then use these regions as constraints to 
predict the ranges for additional observables in $b\to c\tau\nu$ including 
the differential decay distributions $d\Gamma/dq^{2}$;
the ratios $\mathcal{R}(J/\psi)$ and $\mathcal{R}(\Lambda_{c})$;
and the tau-lepton polarisation in $B\to D^{(\star)}\tau\nu$, with
emphasis on the CP violating normal polarisation. Finally we map the allowed regions in $g_s$ and $g_p$ into the parameters of four versions of the 
Yukawa couplings of the general 2HDM-III model. We find that the model is still viable but could be ruled out by a confirmation of a large $\mathcal{R}(J/\psi)$. }

\arxivnumber{1805.04098}

\makeatother

\begin{document}
\maketitle

\section{Introduction}

Amongst the most interesting current results in B physics, the searches
for lepton universality in semileptonic B decays stand out. On the experimental side, hints at deviations
from the standard model (SM) in some of these modes have existed for
several years, with $\bar{B}\to D\tau\nu$ being measured by BaBar
\cite{Lees:2012xj,Lees:2013uzd} and Belle \cite{Huschle:2015rga};
and with $\bar{B}\to D^{\star}\tau\nu$ being measured by BaBar \cite{Lees:2012xj,Lees:2013uzd},
Belle \cite{Huschle:2015rga,Sato:2016svk,Hirose:2016wfn} and LHCb
\cite{Aaij:2015yra,Wormser:2017hsx}. On the theoretical side, many extensions of the SM violate lepton universality whereas the SM does not. The tests involve comparing
semileptonic B decays into tau-leptons to those with muons and electrons through ratios such as, 
\begin{equation}
\mathcal{R}(D^{(*)})=\frac{\Gamma(\bar{B}\to D^{(*)}\tau\bar{\nu})}{\Gamma(\bar{B}\to D^{(*)}l\bar{\nu})},
\end{equation}
where $l$ represents either $e$ or $\mu$. The current values for these quantities 
hint to the existence of new physics, as can be seen when comparing
the current HFLAV averages \cite{Amhis:2016xyh}, 
\begin{eqnarray}
\mathcal{R}(D) & = & 0.407\pm0.039\pm0.024\nonumber \\
\mathcal{R}(D^{\star}) & = & 0.304\pm0.013\pm0.007\ ,\label{exp}
\end{eqnarray}
to the current SM predictions from the lattice for $\mathcal{R}(D)$
\cite{Lattice:2015rga,Na:2015kha} or from a range of models for $\mathcal{R}(D^{\star})$
\cite{Fajfer:2012vx,Bigi:2017jbd}, 
\begin{eqnarray}
\mathcal{R}_{SM}(D) & = & 0.299\pm0.011\nonumber \\
\mathcal{R}_{SM}(D^{\star}) & = & 0.252\pm0.003\ . 
\label{smpred}
\end{eqnarray}
For our new calculations in this paper, we will use the CCQM model for form factors which yields somewhat lower values for these quantities albeit with larger errors, 
$\mathcal{R}_{SM}(D) =  0.27\pm0.03$ and $\mathcal{R}_{SM}(D^{\star})  =  0.24\pm0.02$.

\noindent A related measurement, $B_{c}^{+}\to J/\psi\tau^{+}{\nu}_{\tau}$,
has been reported by LHCb \cite{Aaij:2017tyk} and also hints to
disagreement with the SM, although the errors are too large at present
to reach a definitive conclusion, 
\begin{eqnarray}
\mathcal{R}(J/\psi) & = & \frac{\Gamma(B_{c}^{+}\to J/\psi\tau^{+}{\nu}_{\tau})}{\Gamma(B_{c}^{+}\to J/\psi\mu^{+}{\nu}_{\mu})}\nonumber \\
 & = & 0.71\pm0.17\pm0.18\ .\label{rjpsi}
\end{eqnarray}
Different predictions for the SM arising from different models for
form factors produce a range $0.24$ to $0.28$ \cite{Anisimov:1998uk,Kiselev:2002vz,Ivanov:2006ni,Hernandez:2006gt,Watanabe:2017mip}
which is about 2$\sigma$ lower that the LHCb result. With the CCQM form factors we obtain
\begin{eqnarray}
\mathcal{R}(J/\psi)_{SM}=0.24\pm0.02\ ,
\end{eqnarray}
which we use as the SM prediction in our numerical analysis. 

Not surprisingly, these anomalies have generated enormous interest
in the community. From the experimental side, we expect a measurement
of the corresponding ratio for semileptonic $\Lambda_{b}\to\Lambda_{c}\tau\nu$,
$\mathcal{R}(\Lambda_{c})$ to be reported soon. From the theory side
there have been several proposals for additional observables to be
studied in connection with these modes such as the tau-lepton polarisation
\cite{Tanaka:2012nw,Wu:1997uaa,Lee:2001nw,Chen:2005gr,Chen:2017eby,Iguro:2018qzf}.
In fact, the Belle collaboration has already reported a result for
the longitudinal tau polarisation in $\bar{B}\to D^{*}\tau^{-}\bar{\nu}_{\tau}$
\cite{Hirose:2016wfn} 
\begin{equation}
P_{L}^{\tau}(D^{*})=-0.38\pm0.51_{-0.16}^{+0.21},\label{eq:longPtau}
\end{equation}
a result in agreement with the SM prediction \cite{Tanaka:2012nw}
\begin{equation}
P_{L}^{\tau}(D^{*})_{SM}=-0.497\pm0.013,
\end{equation}
albeit with large uncertainty.

There have also been a large number of theory papers interpreting
these results in the context of specific models, including additional
Higgs doublets, gauge bosons and leptoquarks \cite{Kamenik:2008tj,Tanaka:2010se,Crivellin:2012ye,Celis:2012dk,Deshpande:2012rr,Fajfer:2012jt,Datta:2012qk,Becirevic:2012jf,He:2012zp,Rashed:2012bd,Sakaki:2012ft,Tanaka:2012nw,Ko:2012sv,Bambhaniya:2013wza,Atoui:2013zza,Dutta:2013qaa,Freytsis:2015qca,Boucenna:2016wpr,Boucenna:2016qad,Chiang:2016qov,Zhu:2016xdg,Kim:2016yth,Celis:2016azn,Dutta:2017xmj,Cvetic:2017gkt,Ko:2017lzd,Chen:2017hir,Chen:2017eby,Dutta:2017wpq,He:2017bft,Greljo:2018ogz,Asadi:2018wea,Cline:2015lqp,Crivellin:2015hha}.
One of the first possibilities considered was the 2HDM type II, where
BaBar \cite{Lees:2013uzd} determined it was not possible to simultaneously
fit $\mathcal{R}(D)$ and $\mathcal{R}(D^{\star})$. However, a charged
Higgs with couplings proportional to fermion masses is an obvious
candidate to explain non-universality in semitauonic decays, prompting
consideration of the more general 2HDM-III. Several authors have examined
the flavour phenomenology of the 2HDM-III in the context of the anomalies
mentioned above. Refs.~\cite{Crivellin:2012ye,Tanaka:2012nw,Crivellin:2013wna}
concluded that it is possible to explain $\mathcal{R}(D)$ and $\mathcal{R}(D^{\star})$
in this way after considering existing flavour physics constraints.
More recently, Ref.~\cite{Chen:2017eby,Chen:2018hqy}, add an analysis
of the longitudinal tau-lepton polarisation and forward-backward asymmetries
in $b\to c/u\ \tau\nu$ decays within the 2HDM-III.

In this paper we revisit the $b\to c\tau\nu$ modes in the presence of new (pseudo)-scalar operators to
include several new results. We begin in Section~II with a review of the constraints imposed by the measurements of $\mathcal{R}(D^{(*)})$ and 
the theoretical limit on ${\cal B}(B_{c}\to\tau\nu)$ \cite{Alonso:2016oyd,Akeroyd:2017mhr}. We then use these constraints to obtain the predicted ranges for $\mathcal{R}(J/\psi)$, the tau polarisation in $B\to D^{(*)}\tau\nu$ decays, the differential decay rates and the ratio $\mathcal{R}(\Lambda_{c})$ in Section~III. We pay particular attention  to the transverse tau polarisation which is $T$-odd \cite{Wu:1997uaa,Lee:2001nw,Golowich:1988ig,Tanaka:1994ay,Hagiwara:2014tsa,Ivanov:2017mrj,Garisto:1994vz} as the 2HDM-III model allows for CP violation and would naturally give rise to this effect. 
We also consider the $d\Gamma/dq^{2}$ distributions \cite{Freytsis:2015qca} in $B\to D^{(*)}\tau\nu$ but find that they offer no discriminating power in this case. They do serve to illustrate the CCQM model for the form factors. In Section~IV we review the basics of the general two Higgs doublet model and the four different parameterizations for its Yukawa couplings. We then map this parameter space into the generic allowed regions obtained in Section~II, finding they are completely accessible to this model. Finally, in Section~V we conclude.

\section{$b\to c\tau\nu$  constraints on new (pseudo)-scalar couplings}

The effective Hamiltonian responsible for $b\to c\tau\nu$ transitions
that results from the SM plus the 2HDM-III can be written in terms
of the SM plus generic scalar operators in the form, 
\begin{equation}
\begin{array}{l}
{\cal H}_{{\rm eff}}=C_{SM}^{cb}{\cal O}_{SM}^{cb}+C_{R}^{cb}{\cal O}_{R}^{cb}+C_{L}^{cb}{\cal O}_{L}^{cb},\end{array}
\label{eq:Heffective}
\end{equation}
where $C_{SM}^{cb}=4G_{F}V_{cb}/\sqrt{2}$ and the operators are given
by 
\begin{equation}
\begin{array}{l}
{\cal O}_{SM}^{cb}=\left(\bar{c}\gamma_{\mu}P_{L}b\right)\left(\bar{\tau}\gamma_{\mu}P_{L}\nu_{\tau}\right),\\
{\cal O}_{R}^{cb}=\left(\bar{c}P_{R}b\right)\left(\bar{\tau}P_{L}\nu_{\tau}\right),\\
{\cal O}_{L}^{cb}=\left(\bar{c}P_{L}b\right)\left(\bar{\tau}P_{L}\nu_{\tau}\right).
\end{array}\label{eq:Oeffective}
\end{equation}
As the existing constraints will apply separately to the scalar and the pseudoscalar
couplings, it is convenient to define 
\begin{eqnarray}
g_{S}\equiv\frac{C_{R}^{cb}+C_{L}^{cb}}{C_{SM}^{cb}},\ g_{P}\equiv\frac{C_{R}^{cb}-C_{L}^{cb}}{C_{SM}^{cb}}.
\end{eqnarray}

The effect of the effective Hamiltonian, Eq.~\ref{eq:Heffective},
on the ratios $\mathcal{R}(D^{(*)})$ is known in the literature \cite{Fajfer:2012vx,Crivellin:2012ye,Crivellin:2013wna}
and can be written as ratios $r_{D^{(*)}}={\cal R}(D^{(*)})/{\cal R}_{SM}(D^{(*)})$,
\begin{eqnarray}
r_{D} & = & 1+1.5\ \textrm{Re}\left(g_{S}\right)+1.0\ \left|g_{S}\right|^{2},\nonumber \\
r_{D^{*}} & = & 1+0.12\ \textrm{Re}\left(g_{P}\right)+0.05\ \left|g_{P}\right|^{2}.\label{eq:rD*}
\end{eqnarray}
A few remarks are in order. First, Refs.~\cite{Fajfer:2012jt,Crivellin:2013wna} observe that the coefficient of $|g_{S}|^{2}$ can be changed from 1.0
to 1.5 to  approximate some detector effects in BaBar. As we use the HFLAV average value for $r_{D}$ from both BaBar and Belle results, we will not include this correction in our numerics. 
Second, the CCQM model we use for the form factors leads to the slightly different expression  $r_{D^*}=1+0.1\ \textrm{Re}(g_{P})+0.03\ |g_{P}|^2$, but with larger theoretical errors. We will discuss the effect of this below.

It is also known that there are values of $C_{L}^{cb}$ and $C_{R}^{cb}$
that can explain both of these ratios, and that the possible solutions
become tightly constrained when one also requires that ${\mathcal{B}}(B_{c}\to\tau\nu)\leq30\%$
\cite{Alonso:2016oyd}, which for NP given by scalar operators implies
that the ratio 
\begin{eqnarray}
\frac{{\cal B}(B_{c}\to\tau\nu)}{{\cal B}(B_{c}\to\tau\nu)_{SM}}=\left|1+\frac{m_{B_{c}}^{2}}{m_{\tau}(m_{b}+m_{c})}g_{P}\right|^{2}
\end{eqnarray}
be smaller than around 14.6. An even tighter constraint, by a factor of three, is advocated in Ref.~\cite{Akeroyd:2017mhr}. 

\begin{figure}[!h]
\includegraphics[scale=0.4]{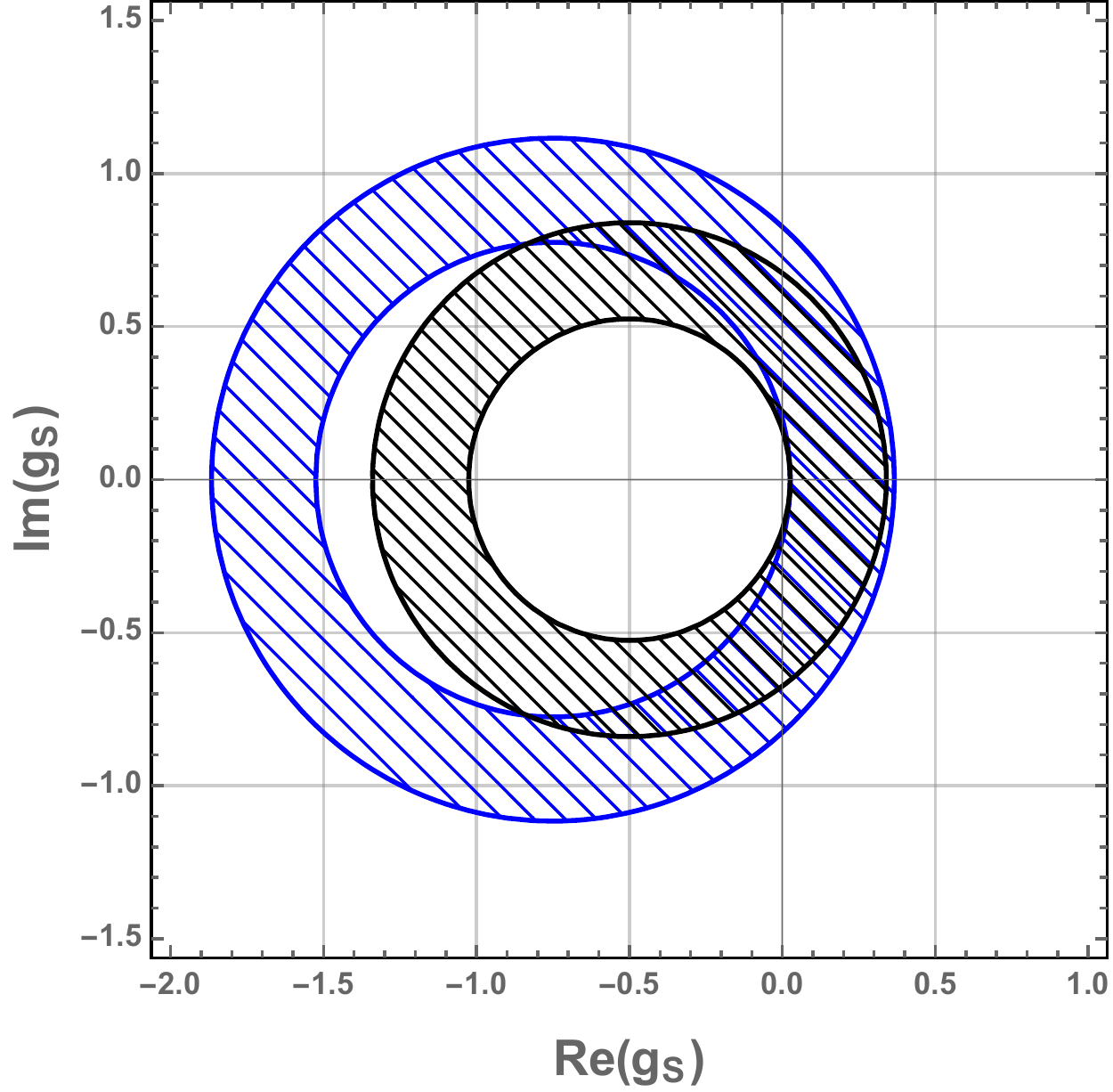}\includegraphics[scale=0.4]{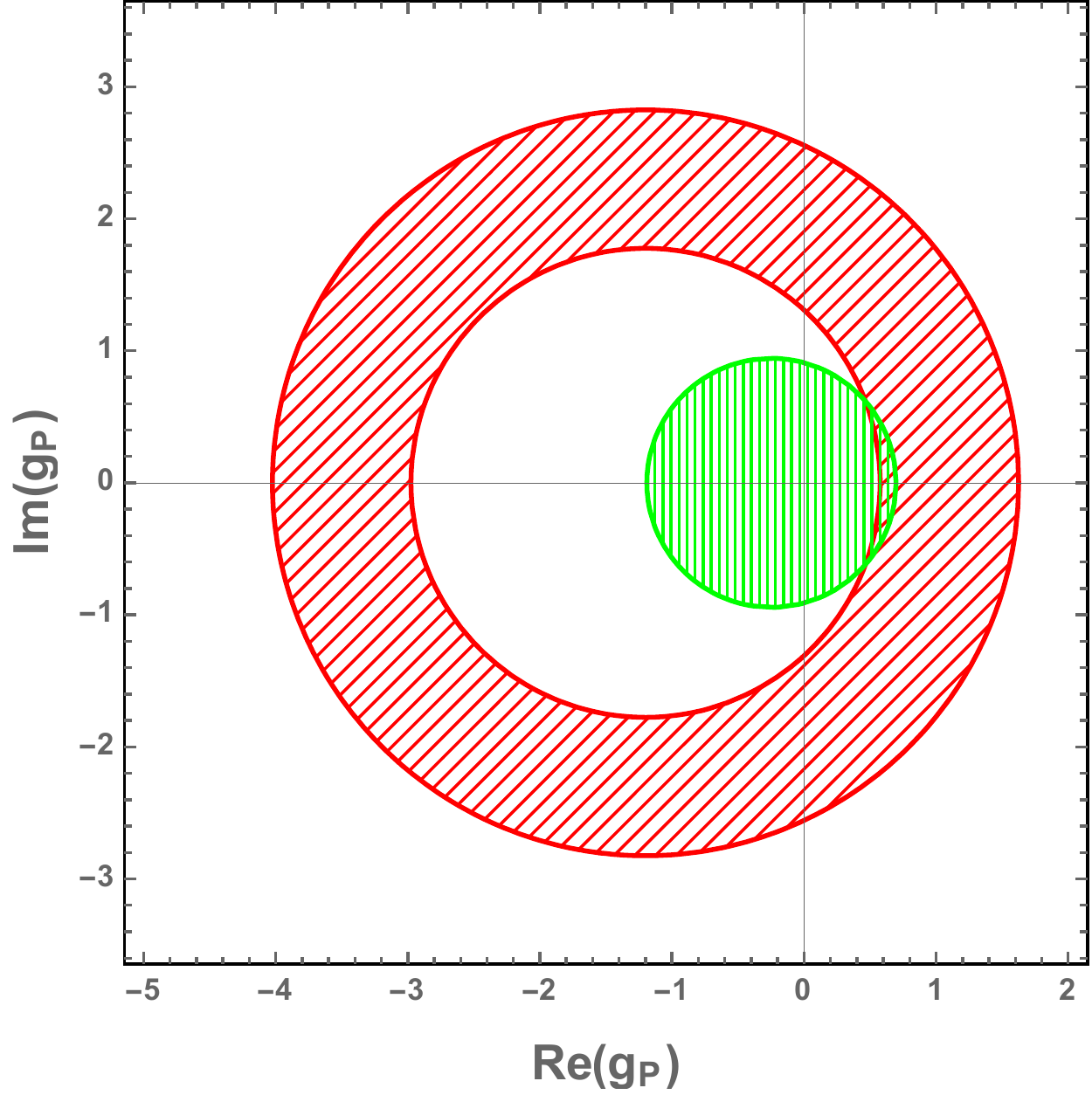}\includegraphics[scale=0.4]{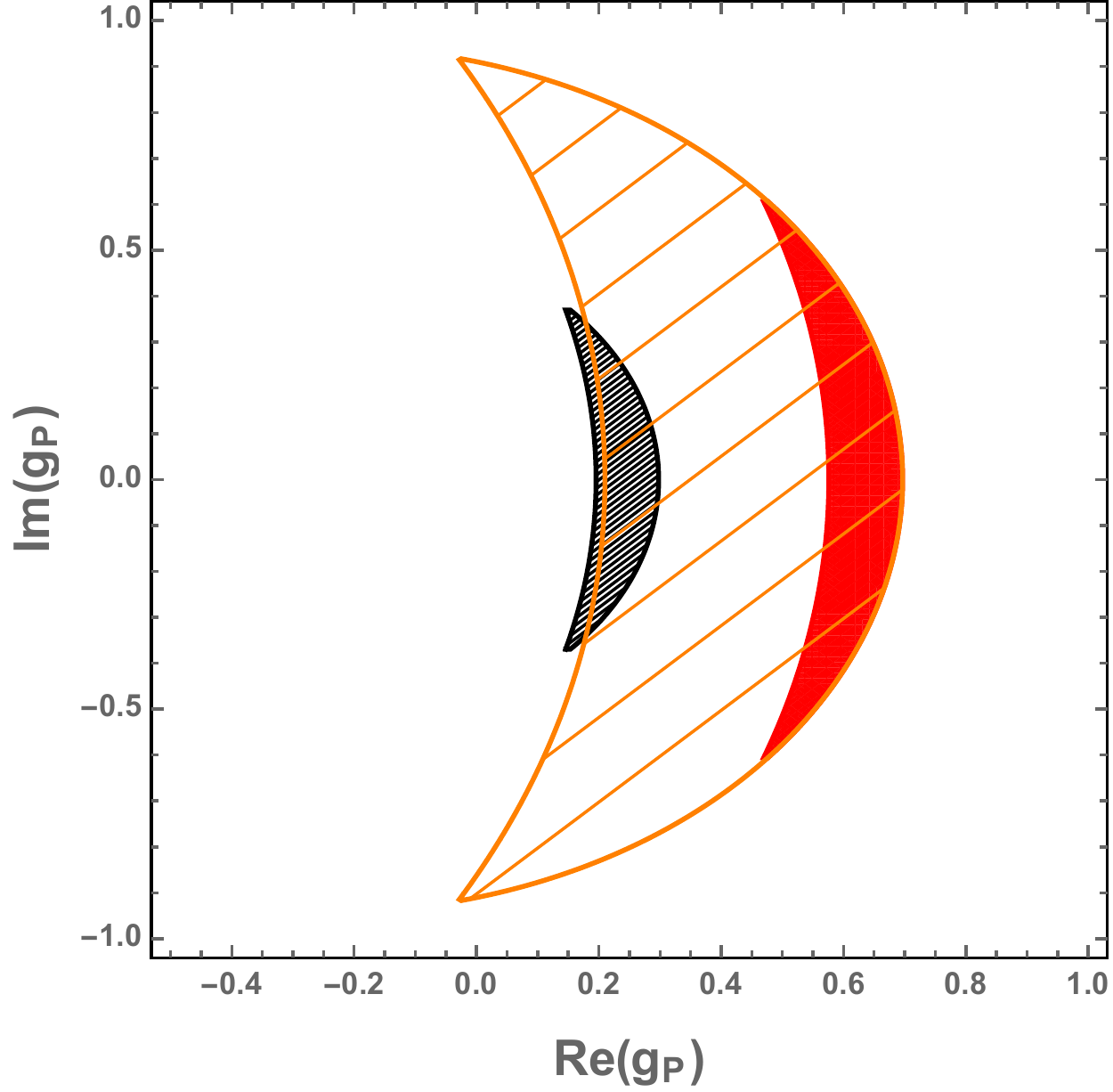}
\caption{Left panel: the blue ring indicates the region allowed by  ${\cal R}(D)$ at the $2\sigma$ level and the black ring shows how this region is modified following the 
prescription of  Refs.~\cite{Fajfer:2012jt,Crivellin:2013wna} described in the text. Center panel: the red ring indicates the region allowed by  ${\cal R}(D^*)$ at the $2\sigma$ level 
and the green circle results in ${\mathcal{B}}(B_{c}\to\tau\nu)\leq30\%$. The small crescent region where these two intersect is the constraint of $g_P$ that we use for our predictions. This region is magnified as the red crescent on the right panel where it is also compared with the larger orange region which uses the CCQM form factors for ${\cal R}(D^*)$, and with the black region which shows the intersection between ${\cal R}(D^*)$ at the $3\sigma$ level and ${\mathcal{B}}(B_{c}\to\tau\nu)\leq10\%$.}
\label{f:known} 
\end{figure}

We summarize these results in Figure~\ref{f:known}. On the left panel we consider the constraint on $g_S$ which arises solely from satisfying ${\cal R}(D)$ at the $2\sigma$ level and appears as the blue ring. The black ring shows the effect of approximating the BaBar detector effects as suggested by Refs.~\cite{Fajfer:2012jt,Crivellin:2013wna}. The central panel shows the constraints on $g_P$: the red ring arising from satisfying $r_{D^*}$ at the $2\sigma$ level and the green circle from ${\mathcal{B}}(B_{c}\to\tau\nu)\leq30\%$. The small combined allowed region shows the tension between these two requirements. On the right panel we illustrate these combined constraints on $g_P$ as the red crescent shape. If one adopts the condition ${\mathcal{B}}(B_{c}\to\tau\nu)\leq10\%$ \cite{Akeroyd:2017mhr} instead, there is no allowed region that also satisfies $r_{D^*}$ at the $2\sigma$ level, but there is one at the $3\sigma$ level and we show this in black. As mentioned above, the expression for $r_{D^*}$ with the CCQM form factors is slightly different but with larger errors which allow a larger overlap with ${\mathcal{B}}(B_{c}\to\tau\nu)\leq30\%$ and this is shown as the orange crescent. For our predictions in the next section we will use the blue ring in the left panel and the red crescent in the right panel. Some, but not all, of these results have appeared before in the literature. For example Refs.~\cite{Ivanov:2016qtw,Shivashankara:2015cta} do not include a constraint from  ${\mathcal{B}}(B_{c}\to\tau\nu)$ in their results.

\section{Predictions}

\subsection{Differential decay distributions for $B\to D^{(\star)}\tau\nu$.}

\begin{figure}[!h]
\includegraphics[scale=0.6]{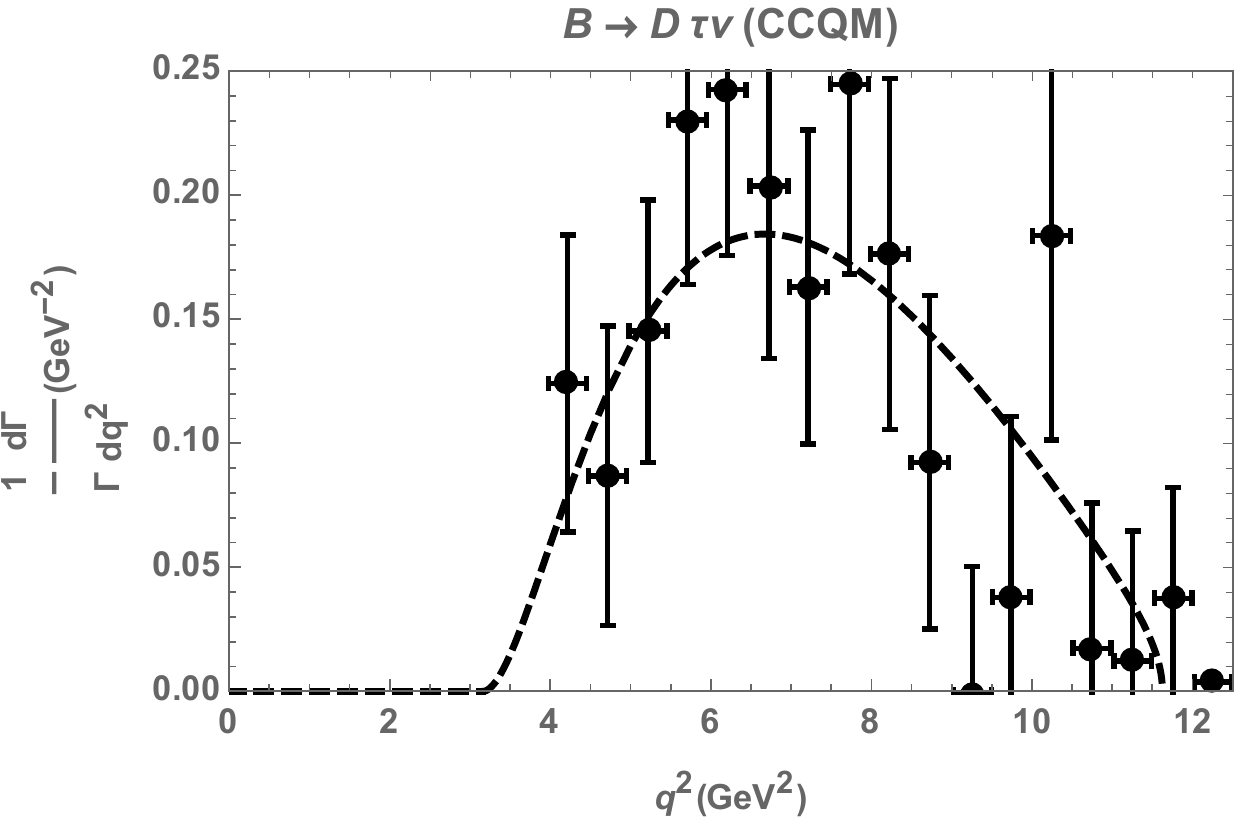}\includegraphics[scale=0.6]{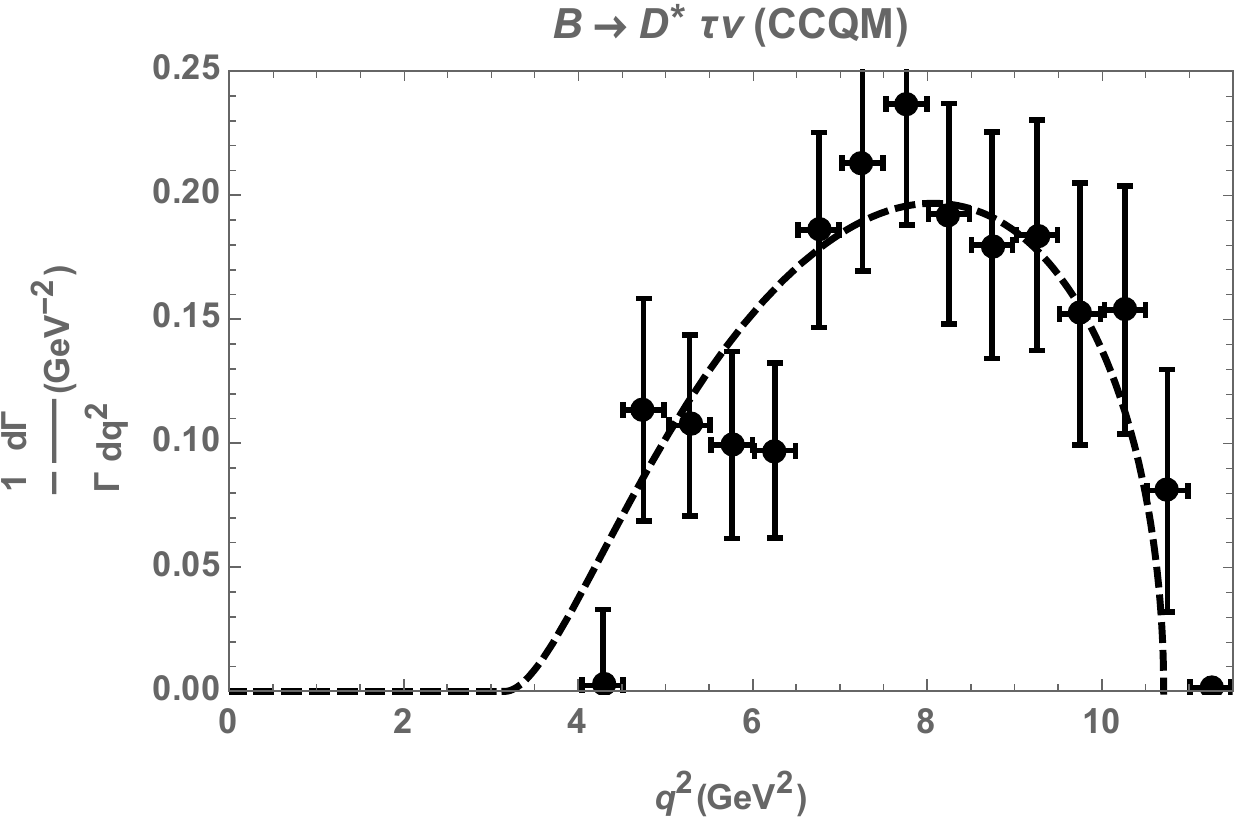}
\caption{Normalised distributions $d\Gamma(B\to D\tau\nu)/(\Gamma dq^{2})$
(left) and $d\Gamma(B\to D^{\star}\tau\nu)/(\Gamma dq^{2})$ (right)
as measured by BaBar \cite{Lees:2013uzd} compared to the predictions
in the SM with form factors from the CCQM of Ref.~\cite{Ivanov:2016qtw}.}
\label{fig:diffBD} 
\end{figure}

In Figure~\ref{fig:diffBD} we compare the distributions $d\Gamma/dq^{2}$
for $B\to D^{(\star)}\tau\nu$ using the CCQM form factors with parameter
values from Ref.~\cite{Ivanov:2016qtw}. The results indicate that
the predicted spectrum is in good agreement with the measurements
within the CCQM uncertainties (which the authors of Ref.~\cite{Ivanov:2016qtw}
estimate at about 10\%). The modifications to these predictions from
$g_{P}$ and $g_{S}$ as constrained above are indistinguishable from
the SM within this level of accuracy.

\subsection{$\mathcal{R}(J/\psi)$}

\begin{figure}[!h]
\center{\includegraphics[scale=0.4]{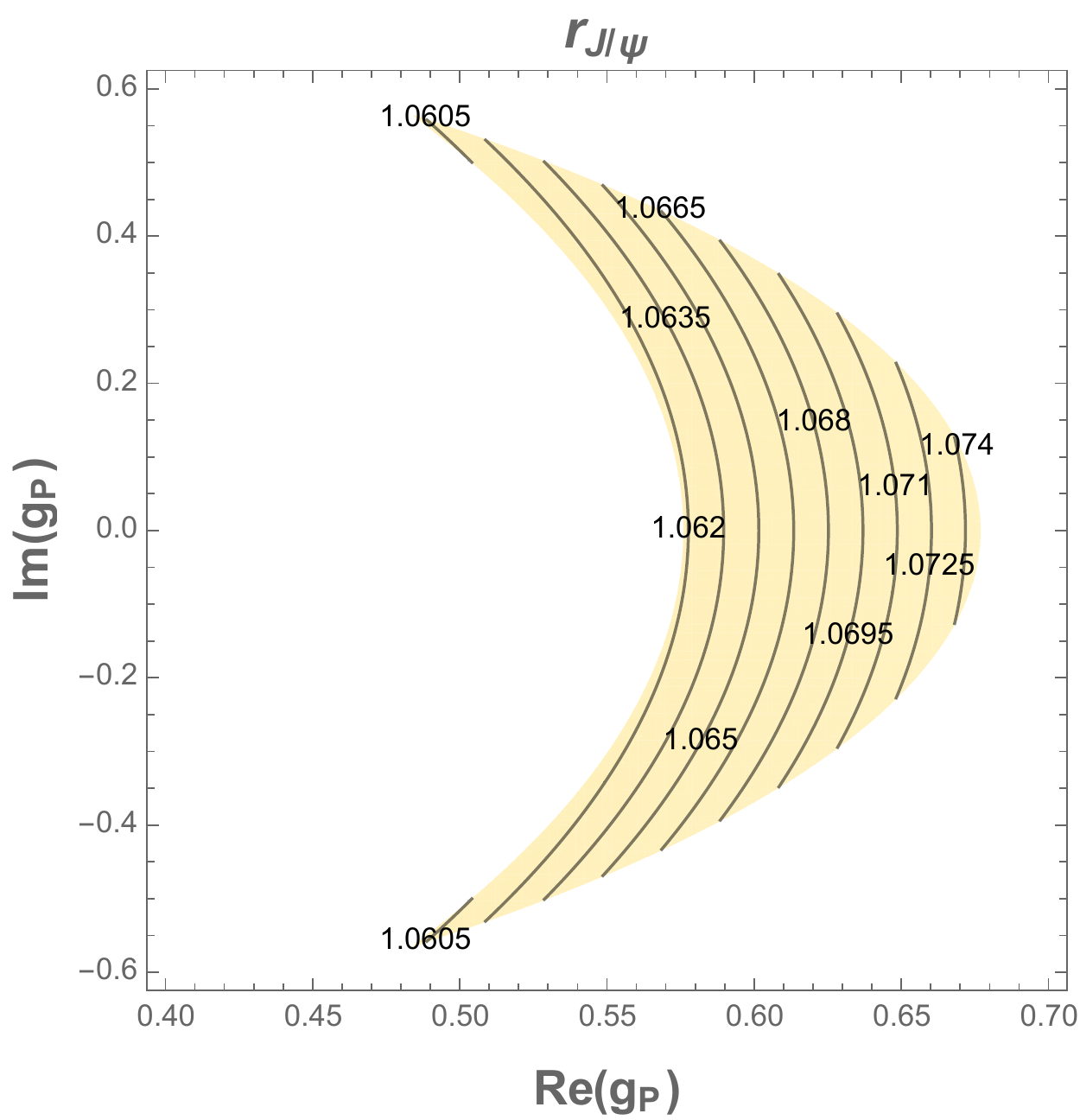}}
\caption{ Predictions for $r_{J/\psi}$ compatible with the measured $\mathcal{R}(D^{\star})$
at $2\sigma$ as well as ${\mathcal{B}}(B_{c}\to\tau\nu)\leq30\%$.}
\label{f:rjpsi} 
\end{figure}

As already mentioned, there is also a more recent measurement of $\mathcal{R}(J/\psi)$ given 
in Eq.~\ref{rjpsi}, which can be used as an additional test of the
model. Using the form factors shown in the appendix with CCQM values
from Ref.~\cite{Tran:2018kuv}, this can be written in terms of
generic scalar coefficients as 
\begin{eqnarray}
r_{J/\psi} & \equiv & \frac{\mathcal{R}(J/\psi)}{\mathcal{R}(J/\psi)_{SM}}\approx1+0.09\ \textrm{Re}\left(g_{P}\right)+0.03\ \left|g_{P}\right|^{2}.
\end{eqnarray}
Note that this result is almost identical to that for $r_{D^{\star}}$
when the CCQM form factors are used for that case as well. The differential
distribution $d\Gamma/dq^{2}$ for $B_{c}\to J/\psi\tau\nu$ receives
tiny corrections from $g_{S,P}$ as constrained above, making it indistinguishable
form the SM one. In Figure~\ref{f:rjpsi} we show the prediction
for $\mathcal{R}(J/\psi)$ that is consistent with the measured $\mathcal{R}(D^{\star})$
at $2\sigma$ as well as ${\mathcal{B}}(B_{c}\to\tau\nu)\leq30\%$.
The largest prediction ($\sim1.075$) is about $1.5\sigma$ away from
the LHCb measurement thanks to its present large uncertainty, which in terms of this ratio is 
$r_{J/\psi} =2.5\pm1.0$. A confirmation of a large value for $r_{J/\psi}$ can potentially rule out (pseudo)-scalar explanations of these anomalies.

\subsection{Polarisations}

In general, we can define normal, longitudinal and transverse polarisations
of the $\tau$ lepton as a function of $q^{2}$ in terms of the vectors
\cite{Ivanov:2017mrj}, 
\begin{eqnarray}
\vec{e}_{L}=\frac{\vec{p}_{\tau}}{|\vec{p}_{\tau}|},\ \vec{e}_{N}=\frac{\vec{p_{\tau}}\times\vec{p}_{D^{(\star)}}}{|\vec{p_{\tau}}\times\vec{p}_{D^{(\star)}}|},\ \vec{e}_{T}=\vec{e}_{N}\times\vec{e}_{L}.\label{eq:Polarization}
\end{eqnarray}
Of particular interest is the normal polarisation, $P_{N}^{\tau}(D^{(\star)})$,
which is generated by CP violating phases that arise from extended
scalar sectors or Yukawa flavour changing couplings\footnote{In the absence of absorptive phases as discussed in the literature.}.
This observable is very small in the SM, where it can only arise due
to unitarity phases in electroweak loop corrections \cite{Wu:1997uaa,Lee:2001nw,Golowich:1988ig,Tanaka:1994ay,Hagiwara:2014tsa,Ivanov:2017mrj,Garisto:1994vz}.

With the numerical CCQM form factors of Ref.~\cite{Ivanov:2017mrj},
we find that new (pseudo)-scalar complex couplings lead to
\begin{eqnarray}
P_{L}^{\tau}(D)\approx\frac{0.33+1.47\ {\rm Re}(g_{S})+0.98|g_{S}|^{2}}{1+\ 1.47{\rm Re}(g_{S})+0.98|g_{S}|^{2}}, &  & P_{L}^{\tau}(D^{\star})\approx\frac{-0.5+0.1\ {\rm Re}(g_{P})+0.03|g_{P}|^{2}}{1+0.1\ {\rm Re}(g_{P})+0.03|g_{P}|^{2}}\nonumber \\
P_{T}^{\tau}(D)\approx\frac{0.84+1.01\ {\rm Re}(g_{S})}{1+\ 1.47{\rm Re}(g_{S})+0.98|g_{S}|^{2}}, &  & P_{T}^{\tau}(D^{\star})\approx\frac{0.46+0.18\ {\rm Re}(g_{P})}{1+0.1\ {\rm Re}(g_{P})+0.03|g_{P}|^{2}} \nonumber \\
P_{N}^{\tau}(D)\approx\frac{-1.01\ {\rm Im}(g_{S})}{1+\ 1.47{\rm Re}(g_{S})+0.98|g_{S}|^{2}}, &  & P_{N}^{\tau}(D^{\star})\approx \frac{-0.18\ {\rm Im}(g_{P})}{1+0.1\ {\rm Re}(g_{P})+0.03|g_{P}|^{2}}
\label{polshiggs}
\end{eqnarray}

\begin{figure}[!h]
\includegraphics[scale=0.4]{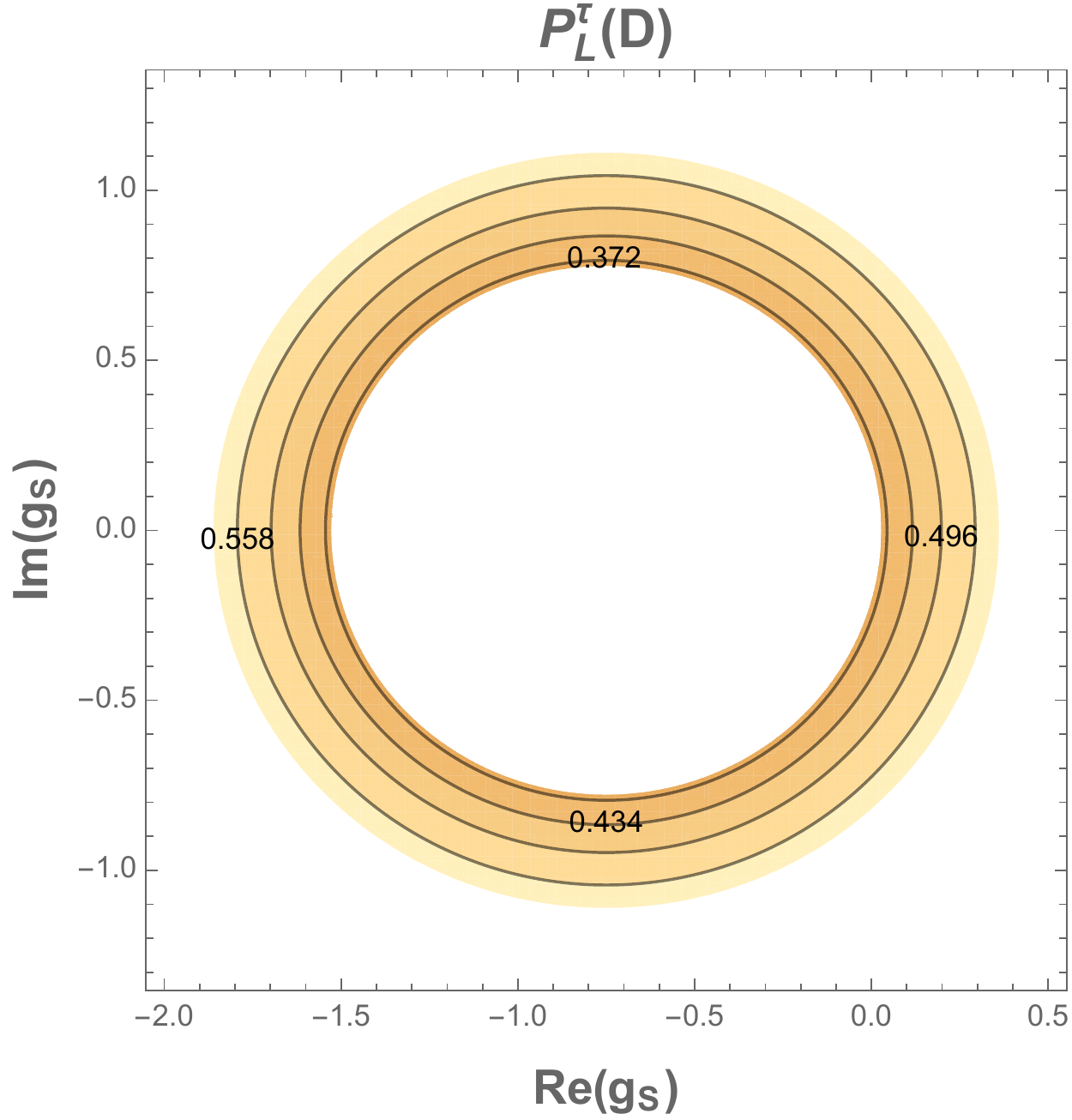}\includegraphics[scale=0.4]{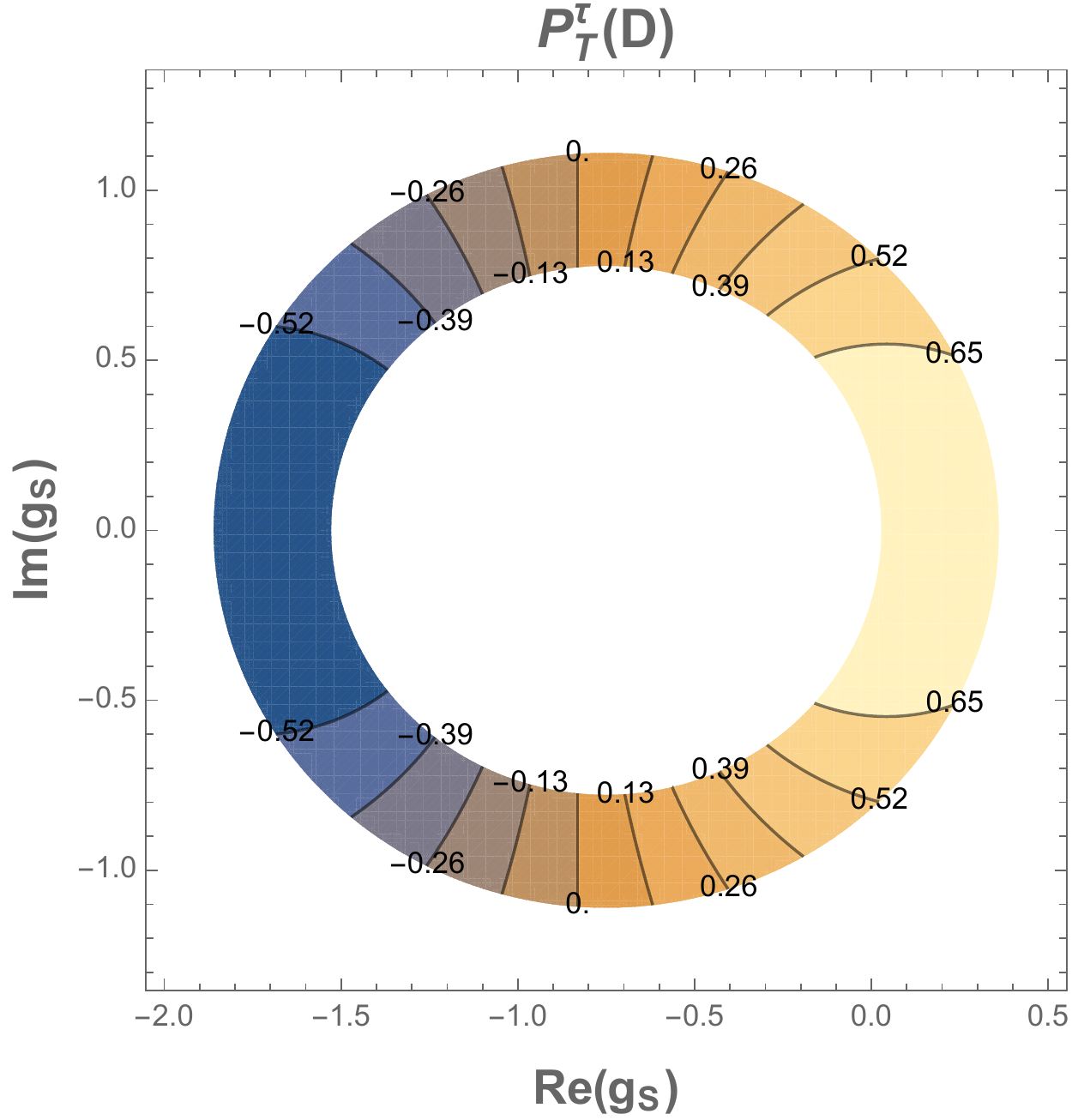}\includegraphics[scale=0.4]{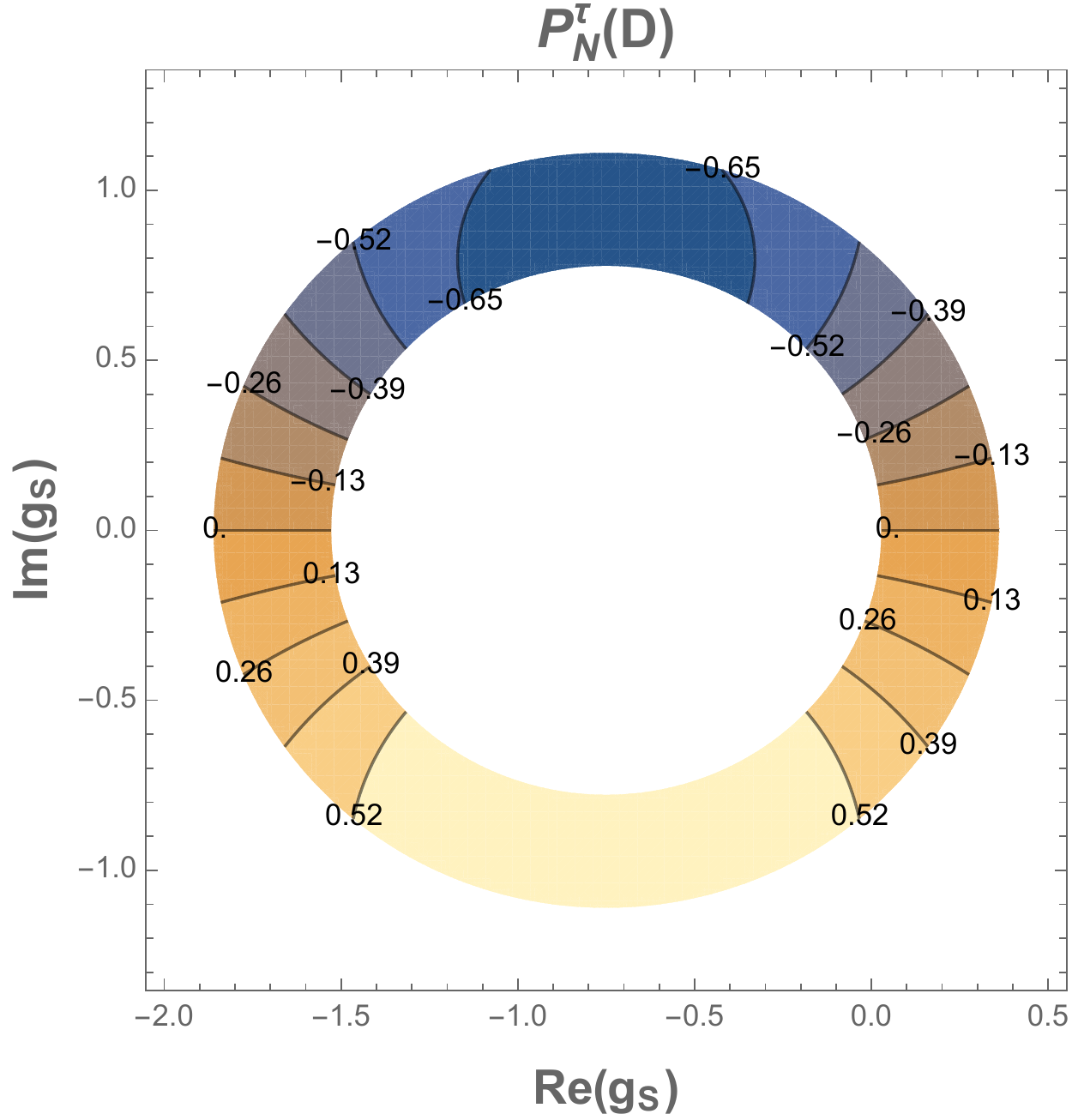}
\caption{Average tauonic polarisations $P_{L,N,T}^{\tau}(D)$ for the allowed
parameter space.}
\label{fig:Average-PD} 
\end{figure}

\begin{figure}[!h]
\includegraphics[scale=0.4]{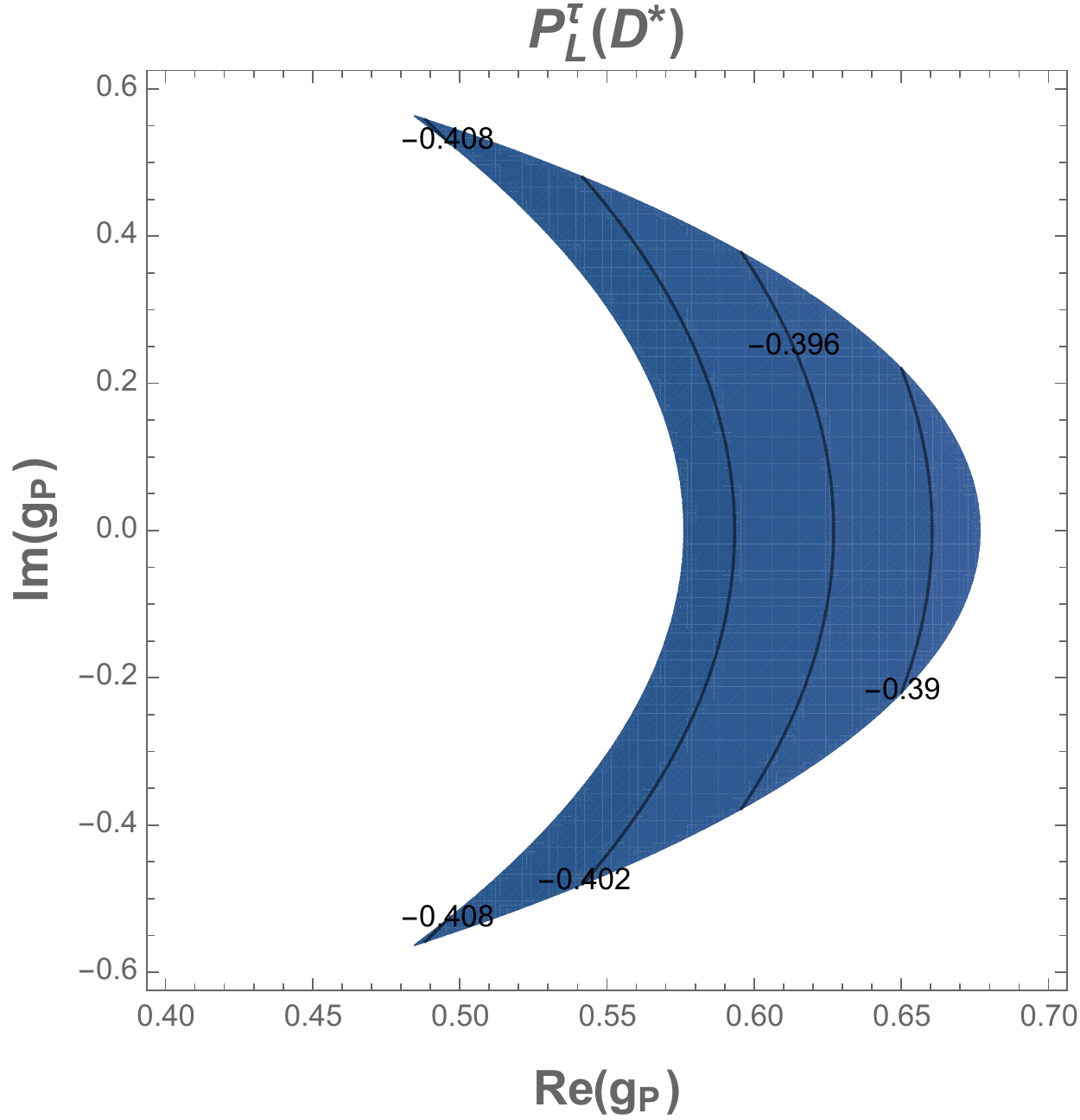}\includegraphics[scale=0.4]{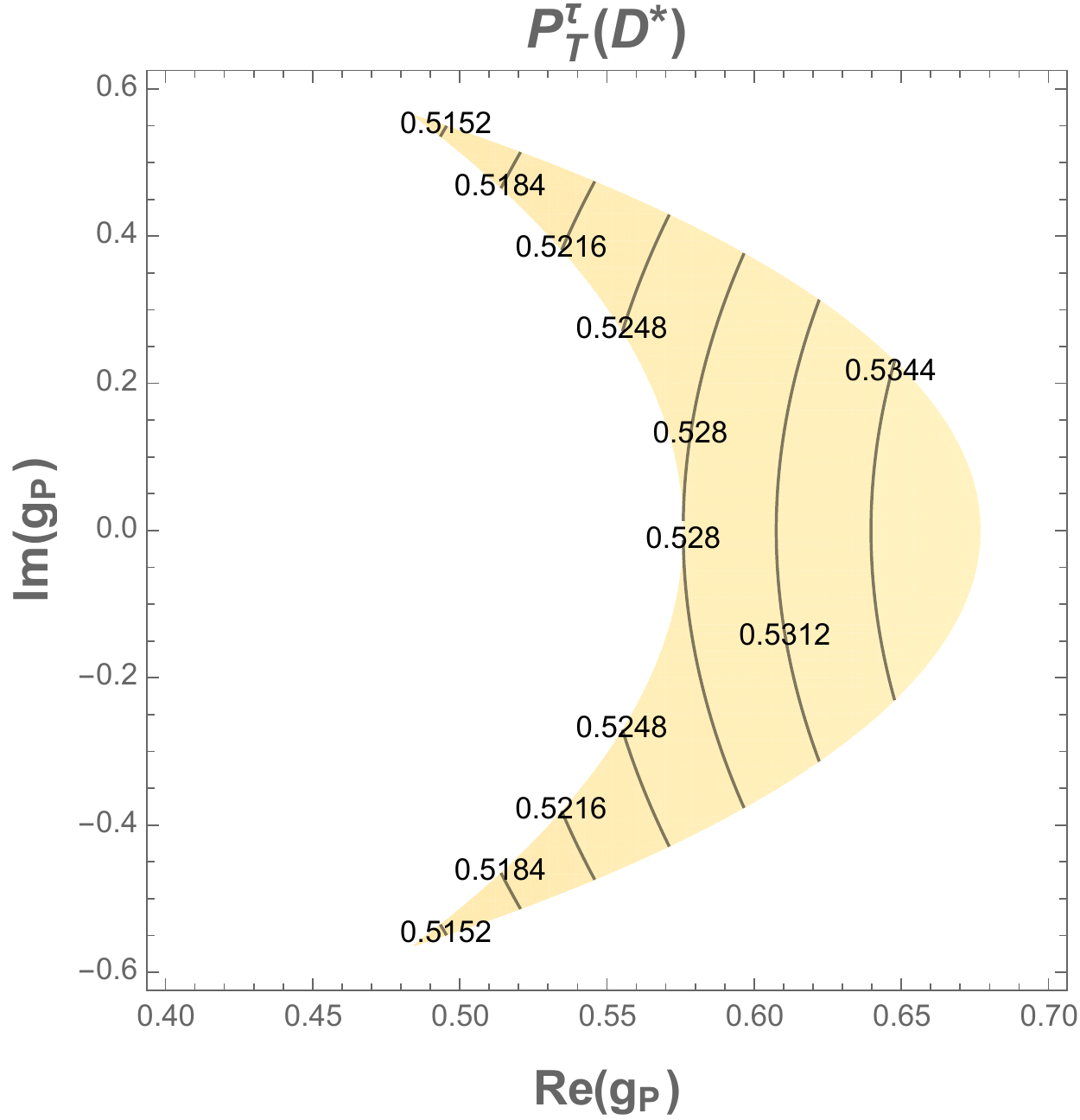}\includegraphics[scale=0.4]{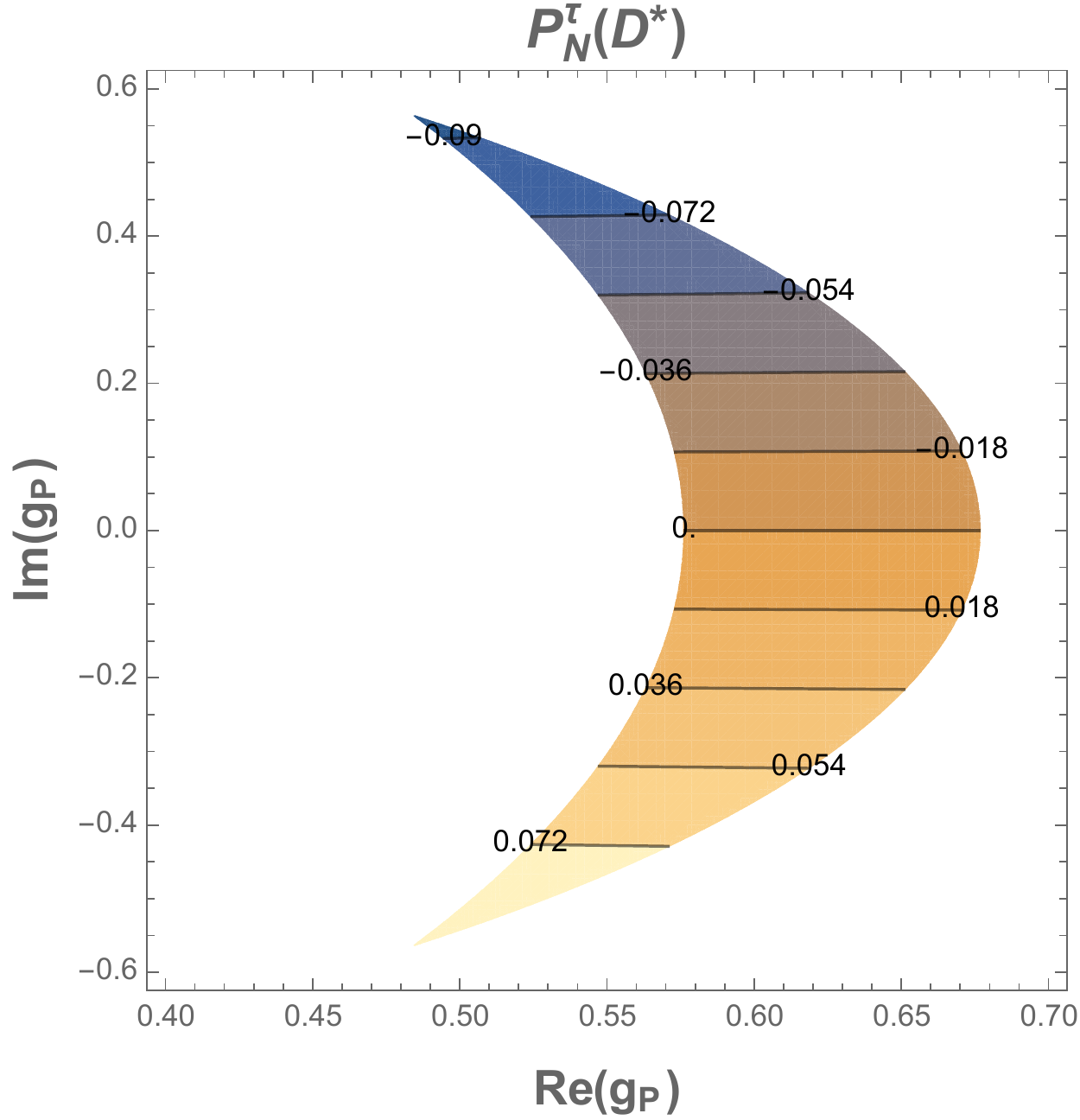}
\caption{Average tauonic polarisations $P_{L,N,T}^{\tau}(D^{*})$ for the allowed
parameter space. }
\label{fig:Average-PD*} 
\end{figure}

Figures \ref{fig:Average-PD}-\ref{fig:Average-PD*} show Eq.(\ref{polshiggs})
in the allowed parameter regions obtained above. In particular we
see that $P_{L}^{\tau}(D^{\star})$ as measured by Belle \cite{Hirose:2016wfn}
is consistent with all the predictions given the current large uncertainty.
The figures also indicate that a large CP violating $P_{N}^{\tau}(D)$ polarisation is
possible.

\subsection{$\Lambda_{b}\rightarrow\Lambda_{c}l\bar{\nu}$ decays}

As mentioned in the introduction, there is one more ratio in the $b\to c\tau\nu$
family that is expected to be measured soon by LHCb, namely $\mathcal{R}(\Lambda_{c})$.
In terms of the CCQM form factors we show the differential decay
rate  \cite{Li:2016pdv,Shivashankara:2015cta} in the appendix.

\begin{figure}[!h]
\center{\includegraphics[scale=0.8]{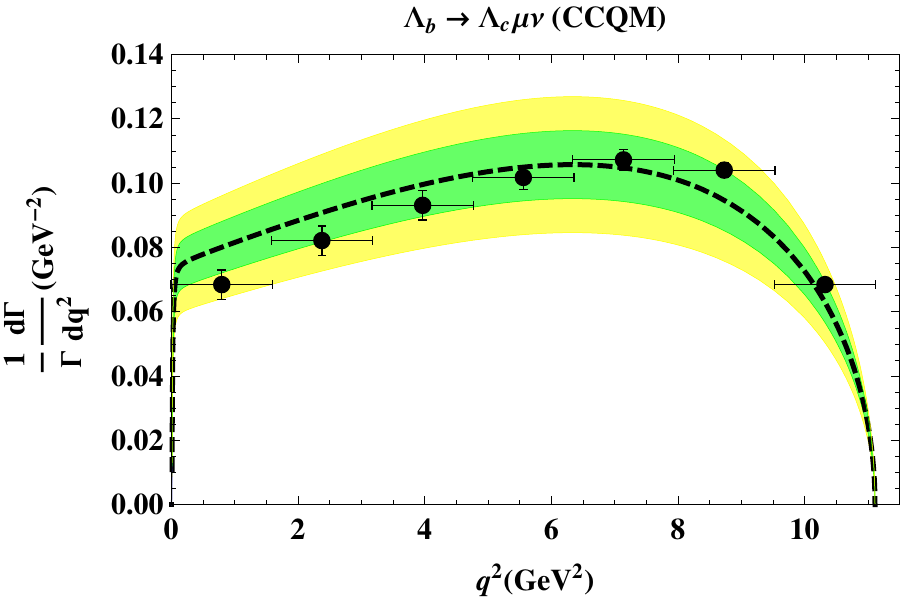}}
\caption{Normalised distribution $d\Gamma(\Lambda_{b}\to\Lambda_{c}\mu\bar{\nu}_{\mu})/(\Gamma dq^{2})$
		 with the CCQM form factors. The points are the LHCb data \cite{Aaij:2017svr}
		and the dashed black line, green and yellow areas mark the central
		CCQM results and a 10\% and 20\% deviation respectively. \label{fig:-spectral-distribution} }
\end{figure}

From the partial decay width Eq.(\ref{eq:lambda decay width}), we
first obtain the $\Lambda_{b}\to\Lambda_{c}\mu\bar{\nu}_{\mu}$ normalised
spectral distribution for the SM and compare it with the one measured
by LHCb \cite{Aaij:2017svr} in  Figure \ref{fig:-spectral-distribution}.
The green and yellow shaded areas indicate the estimated 10\% and
20\% errors in the prediction according to \cite{Ivanov:2017mrj}.
Once again, this figure serves to calibrate the performance of the CCQM form factors
in this case. 

\begin{figure}[!h]
\center{\includegraphics[scale=0.5]{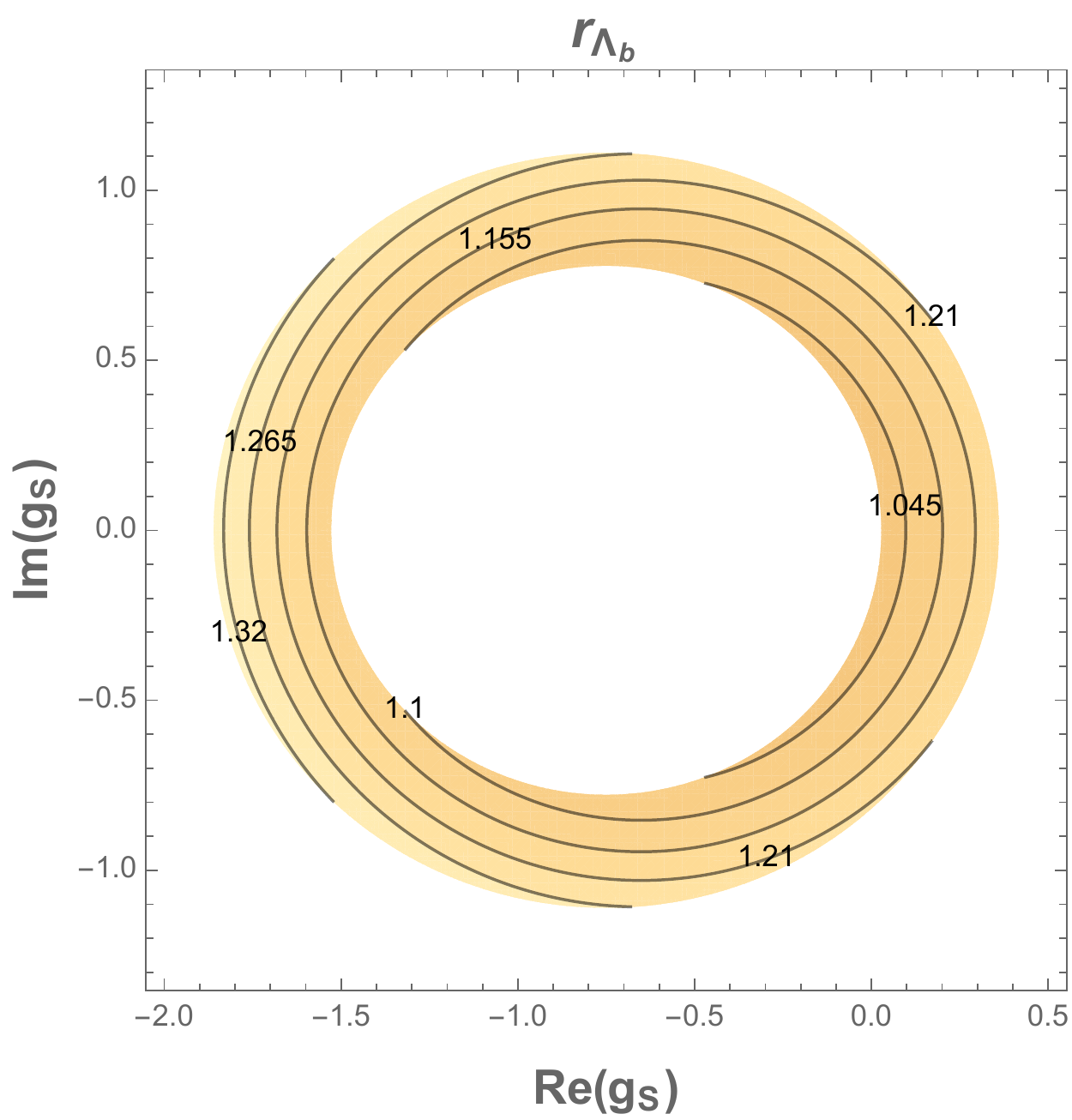}\includegraphics[scale=0.5]{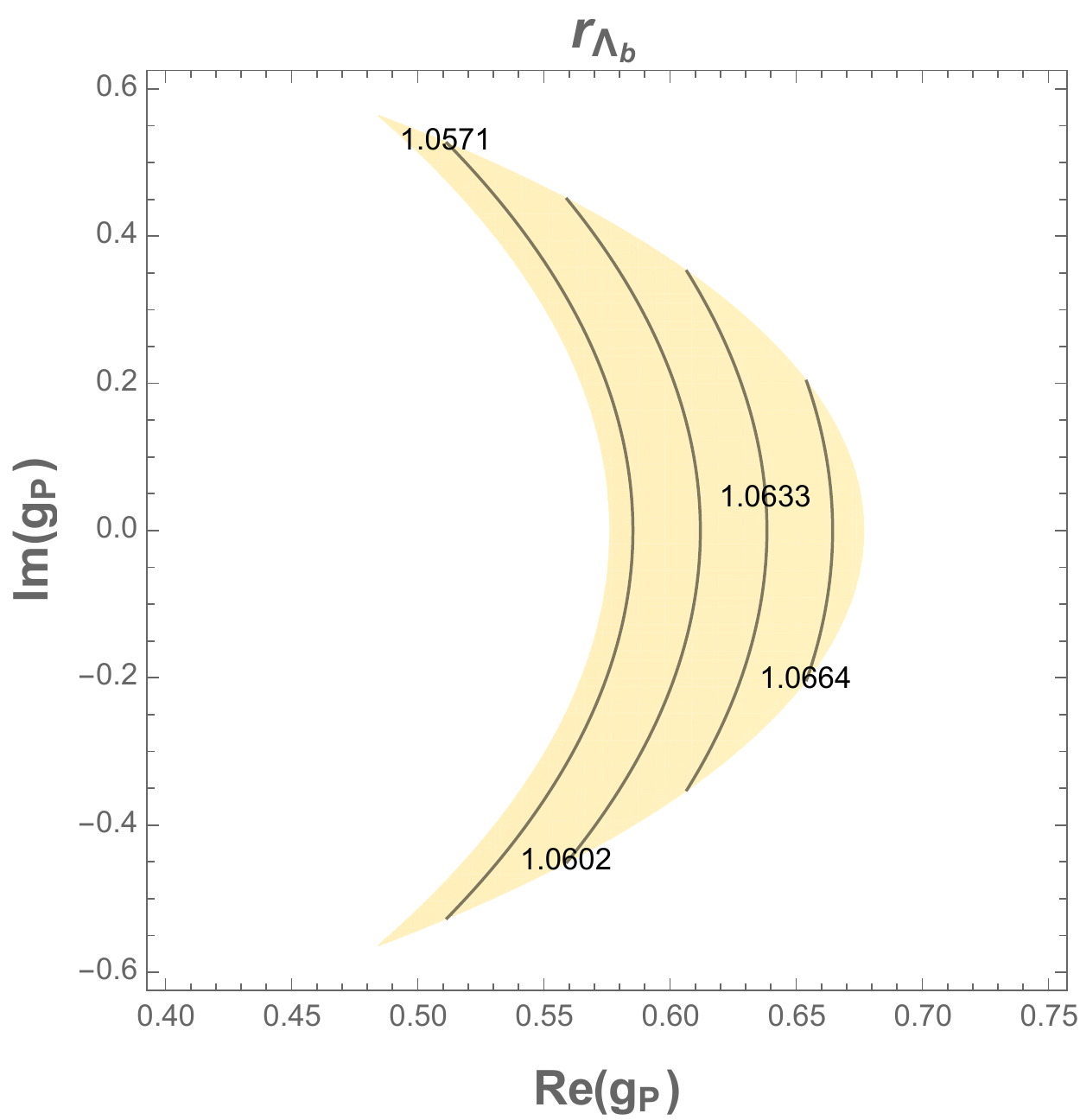}}
\caption{ Predictions for $r_(\Lambda_{c})$ compatible with the measured $\mathcal{R}(D^{\star})$
at $2\sigma$ as well as ${\mathcal{B}}(B_{c}\to\tau\nu)\leq30\%$. The left and right panels show separately the contributions from $g_S$ and $g_P$.}
\label{f:rjlambda} 
\end{figure}

We also find in this case that the spectral distribution with new $g_{S,P}$ couplings constrained as above, cannot differentiate between
the models. We turn to a prediction for $\mathcal{R}(\Lambda_{c})$
which is defined analogously to the previous ratios, 
\begin{eqnarray}
\mathcal{R}(\Lambda_{c})  \equiv  \frac{\Gamma(\Lambda_{b}\to\Lambda_{c}\tau\bar{\nu})}{\Gamma(\Lambda_{b}\to\Lambda_{c}l\bar{\nu})},&&
r(\Lambda_{c}) =\frac{\mathcal{R}(\Lambda_{c}) }{\mathcal{R}(\Lambda_{c})_{SM} }.
\end{eqnarray}
With the form factors in the appendix this leads to 
\begin{eqnarray}
r(\Lambda_{c})\approx\left(1+0.08\text{Re}[g_{P}]+0.43\text{Re}[g_{S}]+0.03|g_{P}|^{2}+0.33|g_{S}|^{2}\right).
\label{eqrl}
\end{eqnarray}
It also leads to $\mathcal{R}(\Lambda_{c})_{SM}=0.295$,  which compares well with other values found in the literature $\mathcal{R}(\Lambda_{c})_{SM}=0.33\pm0.01$
\cite{Li:2016pdv}. Figure \ref{f:rjlambda} shows $\mathcal{R}(\Lambda_{c})$  with new contributions from $g_S$ or $g_P$ in their allowed ranges. As Eq.~\ref{eqrl} shows no interference between $g_S$ and $g_P$, the two new contributions simply add.

\section{General two Higgs doublet model}

The most general 2HDM-III, unlike the type I and type II more common
versions, allows flavour changing neutral currents (FCNC) at tree-level
which are then suppressed with family symmetries, minimal flavour
violation, or specific patterns for the Yukawa couplings, for example.
The most general renormaliseable quartic scalar potential is commonly
written as \cite{Wu:1994ja}, 
\begin{eqnarray}
V(\Phi_{1},\Phi_{2}) & = & \mu_{1}^{2}(\Phi_{1}^{\dag}\Phi_{1})+\mu_{2}^{2}(\Phi_{2}^{\dag}\Phi_{2})-\left(\mu_{12}^{2}(\Phi_{1}^{\dag}\Phi_{2})+h.c\right)\nonumber \\
 & + & \frac{1}{2}\lambda_{1}(\Phi_{1}^{\dag}\Phi_{1})^{2}+\frac{1}{2}\lambda_{2}(\Phi_{2}^{\dag}\Phi_{2})^{2}+\lambda_{3}(\Phi_{1}^{\dag}\Phi_{1})(\Phi_{2}^{\dag}\,\Phi_{2})+\lambda_{4}(\Phi_{1}^{\dag}\Phi_{2})(\Phi_{2}^{\dag}\Phi_{1})\nonumber \\
 & + & \left(\frac{1}{2}\lambda_{5}(\Phi_{1}^{\dag}\Phi_{2})^{2}+\left(\lambda_{6}(\Phi_{1}^{\dag}\Phi_{1})+\lambda_{7}(\Phi_{2}^{\dag}\Phi_{2})\right)(\Phi_{1}^{\dag}\Phi_{2})+{\rm ~h.c.}\right)\label{eq:Higgs potential}
\end{eqnarray}
where the two scalar doublets are 
\begin{equation}
\ensuremath{\Phi_{i}=\left(\begin{array}{c}
\phi_{i}^{+}\\
\frac{1}{\sqrt{2}}(\upsilon_{i}+\xi_{i}+i\zeta_{i})
\end{array}\right)}.
\end{equation}
Discrete symmetries in 2HDM type I and II force the parameters $\mu_{12}$
and $\lambda_{6,7}$ to vanish. The charged Higgs bosons that appear
in the mass eigenstate basis correspond to the combinations 
\begin{eqnarray}
H^{\pm}=-s_{\beta}\phi_{1}^{\pm}+c_{\beta}\phi_{2}^{\pm}\label{defbeta}
\end{eqnarray}
with the rotation angle given by $\tan\beta=t_{\beta}=\frac{s_{\beta}}{c_{\beta}}=\frac{\upsilon_{2}}{\upsilon_{1}}$
and $\upsilon_{2}^{2}+\upsilon_{1}^{2}=\upsilon^{2}$ with $\upsilon=246$~GeV.
There are three neutral scalars that are not CP eigenstates as the
parameters $\mu_{12}$ and $\lambda_{6,7}$ can be complex and violate
CP. These, however, will not play any role in our discussion beyond
the occasional use of existing constraints on the mixing amongst the
neutral scalars.

The most general Yukawa Lagrangian in the 2HDM-III without discrete
symmetries is given by 
\begin{eqnarray}
 &  & \mathcal{L}_{Y}=-\left(\overline{Q}_{L}Y_{1}^{u}u_{R}\widetilde{\Phi}_{1}+\overline{Q}_{L}Y_{2}^{u}u_{R}\widetilde{\Phi}_{2}+\overline{Q}_{L}Y_{1}^{d}d_{R}\Phi_{1}\right.\nonumber \\
 &  & +\left.\overline{Q}_{L}Y_{2}^{d}d_{R}\Phi_{2}+\overline{L}_{L}Y_{1}^{l}l_{R}\Phi_{1}+\overline{L}_{L}Y_{2}^{l}l_{R}\Phi_{2}\right)+{\rm ~h.c.}\\
\nonumber 
\end{eqnarray}
where $\widetilde{\Phi}_{1,2}=i\sigma_{2}\Phi_{1,2}^{*}$, $Q_{L}$
and $L_{L}$ denote the left-handed quark and lepton doublets, $u_{R}$,
$d_{R}$ and $l_{R}$ the right-handed quark and lepton singlets and
$Y_{1,2}^{u,d,l}$ denote the $(3\times3)$ Yukawa matrices.

After spontaneous EWSB and in the fermion mass basis, the charged
Higgs couplings to fermions can be written as: 
\begin{eqnarray}
{\cal L}^{\bar{f}_{i}f_{j}\phi} & = & -\frac{g}{2\sqrt{2}M_{W}}\Bigg[\sum_{l=1}^{3}\bar{u}_{i}\Bigg[(V_{{\rm CKM}})_{il}\left(X\,m_{d_{l}}\,\delta_{lj}-\frac{f(X)}{\sqrt{2}}\,\sqrt{m_{d_{l}}m_{d_{j}}}\,\tilde{\chi}_{lj}^{d}\right)(1+\gamma^{5})\nonumber \\
 &  & +\left(Y\,m_{u_{i}}\,\delta_{il}-\frac{f(Y)}{\sqrt{2}}\,\sqrt{m_{u_{i}}m_{u_{l}}}\,\tilde{\chi}_{il}^{u}\right)(V_{{\rm CKM}})_{lj}(1-\gamma^{5})\Bigg]\,d_{j}\,H^{+}\label{eq:cChargedHiggs}\\
 &  & +\bar{\nu}_{i}\bigg(Z\,m_{l_{i}}\,\delta_{ij}-\frac{f(Z)}{\sqrt{2}}\,\sqrt{m_{l_{i}}m_{l_{j}}}\,\tilde{\chi}_{ij}^{l}\bigg)(1+\gamma^{5}){l}_{j}H^{+}+{\rm ~h.c.}\Bigg]\nonumber 
\end{eqnarray}
where $f(x)=\sqrt{1+x^{2}}$. This form follows the notation of Refs.~\cite{HernandezSanchez:2012eg,DiazCruz:2004pj,DiazCruz:2004tr}
in which the first term in each line in Eq.~\ref{eq:cChargedHiggs}
is the coupling in one of the four 2HDM without FCNC and the second
term is a flavour changing correction that makes it a type III model.
Furthermore, the Cheng-Sher ansatz \cite{Cheng:1987rs} has been
implemented to control the size of the FCNC, but also allowing a CP
violating phase: 
\begin{equation}
\left[\tilde{Y}^{q,l}\right]_{ij}=\frac{\sqrt{m_{i}^{q,l}m_{j}^{q,l}}}{\upsilon}\,\left[\tilde{\chi}^{q,l}\right]_{ij}\label{sher}
\end{equation}
with $\tilde{Y}^{f}=V_{fL}^{\dagger}Y^{f}V_{fR}$. The additional
parameters that occur as a consequence of allowing flavour changing
couplings are $\tilde{\chi}^{q.l}_{ij}$.

The parameters $X$, $Y$ and $Z$ given in Table \ref{tab:Parameters}
are the ones that occur in each of the four types of 2HDM with natural
flavour conservation. 
\begin{table}
\centering %
\begin{tabular}{|c|c|c|c|}
\hline 
2HDM-III  & $X$  & $Y$  & $Z$ \tabularnewline
\hline 
Model I  & $-\cot\beta$  & $\cot\beta$  & $-\cot\beta$ \tabularnewline
Model II  & $\tan\beta$  & $\cot\beta$  & $\tan\beta$ \tabularnewline
Model X  & $-\cot\beta$  & $\cot\beta$  & $\tan\beta$ \tabularnewline
Model Y  & $\tan\beta$  & $\cot\beta$  & $-\cot\beta$ \tabularnewline
\hline 
\end{tabular}\caption{Parameters $X$, $Y$ and $Z$ defined in the Yukawa interactions
of Eq.(\ref{eq:cChargedHiggs}) for four versions of the 2HDM-III.\label{tab:Parameters}}
\end{table}

We now turn to the question of the scalar coefficients in Eq.~\ref{eq:Heffective} within the. 
context of the 2HDM-III considered here. Tree-level exchange of the
charged Higgs produces 
\begin{eqnarray}
C_{L}^{cb}=\frac{-1}{m_{H^{\pm}}^{2}}\Gamma_{cb}^{L}\Gamma_{\nu\tau}^{R}, &  & C_{R}^{cb}=\frac{-1}{m_{H^{\pm}}^{2}}\Gamma_{cb}^{R}\Gamma_{\nu\tau}^{R},\label{eq:Wilson R}
\end{eqnarray}
and Eq.(\ref{eq:cChargedHiggs}) implies that, 
\begin{eqnarray}
\Gamma_{ij}^{L} & = & \frac{g}{\sqrt{2}M_{W}}\sum_{l=1}^{3}\left(Y\,m_{u_{i}}\,\delta_{il}-\frac{f(Y)}{\sqrt{2}}\,\sqrt{m_{u_{i}}m_{u_{l}}}\,\tilde{\chi}_{il}^{u}\right)(V_{{\rm CKM}})_{lj},\nonumber \\
\Gamma_{ij}^{R} & = & \frac{g}{\sqrt{2}M_{W}}\sum_{l=1}^{3}(V_{{\rm CKM}})_{il}\left(X\,m_{d_{l}}\,\delta_{lj}-\frac{f(X)}{\sqrt{2}}\,\sqrt{m_{d_{l}}m_{d_{j}}}\,\tilde{\chi}_{lj}^{d}\right),\nonumber \\
\Gamma_{\nu_{i}j}^{R} & = & \frac{g}{\sqrt{2}M_{W}}\sum_{i=1}^{3}\bigg(Z\,m_{l_{i}}\,\delta_{ij}-\frac{f(Z)}{\sqrt{2}}\,\sqrt{m_{l_{i}}m_{l_{j}}}\,\tilde{\chi}_{ij}^{l}\bigg).\label{coupsch}
\end{eqnarray}
Assuming that the parameters $\tilde{\chi}_{i,j}^{u}$ are of the
same order and that the $\tilde{\chi}_{i,j}^{d}$ are also of the
same order, as we expect in the context of the Cheng-Sher ansatz,
the contributions from the heaviest fermions dominate the sums and
Eq.~\ref{coupsch} reduces to 
\begin{eqnarray}
\Gamma_{cb}^{L} \simeq  \frac{g}{\sqrt{2}M_{W}}m_{c}V_{cb}\tilde Y,&&\tilde Y= \left(Y\,-\frac{V_{tb}}{V_{cb}}\frac{f(Y)}{\sqrt{2}}\,\sqrt{\frac{m_{t}}{m_{c}}}\,\tilde{\chi}_{ct}^{u}\right),\nonumber \\
\Gamma_{cb}^{R}  \simeq  \frac{g}{\sqrt{2}M_{W}}m_{b}V_{cb}\tilde X,&&\tilde X= \left(X\,-\frac{V_{cs}}{V_{cb}}\frac{f(X)}{\sqrt{2}}\,\sqrt{\frac{m_{s}}{m_{b}}}\,\tilde{\chi}_{sb}^{d}\right),\nonumber \\
\Gamma_{\nu\tau}^{R}  \simeq  \frac{g}{\sqrt{2}M_{W}}m_{\tau}\tilde Z, &&\tilde Z=\left(Z-\frac{f(Z)}{\sqrt{2}}\,\tilde{\chi}_{\tau\tau}^{l}\right).\label{cimplecou}
\end{eqnarray}
The allowed parameter regions of Figure~\ref{f:known} then imply
constraints on the parameters $m_{H^{\pm}},\,\tan\beta,\,\tilde{\chi}_{ij}^{u,d,l}$
which we discuss next in some detail. The general result is that it is possible to reach the allowed regions in Figure~\ref{f:known} with parameters
of the model. Ref.~\cite{Arbey:2017gmh} finds solutions for generalised models
which can be written in terms of our Eqs.~\ref{cimplecou} with the
factors $\tilde X,\tilde Y,\tilde Z$ being arbitrary parameters, independent of $\tan\beta$. The solutions
they find occur for points with $\tilde X\sim{\cal O}(10)$, $\tilde Y\sim{\cal O}(100)$, $\tilde Z\sim{\cal O}(100)$
and $m_{H}<550$~GeV.

Once we allow for FCNC, all four cases of 2HDM-III can be mapped into the allowed regions in Figure~\ref{f:known}. In
Figures~\ref{f:mapto31}-\ref{f:mapto3y} we illustrate the results
in two dimensional projections of parameter space. In all cases we present three figures. In the first one we consider the plane $\tan\beta-m_{H^\pm}$ and scan over all the real and imaginary parts of the $\tilde{\chi}$ parameters looking for points that satisfy the primary constraints ${\cal R}(D^*)$ at $2\sigma$ and ${\mathcal{B}}(B_{c}\to\tau\nu)\leq30\%$. In the second and third plots we illustrate regions of the parameter space of the $\tilde{\chi}$'s where the constraints are satisfied, in particular we specifically show solutions in the vicinity of $g_s=-0.5+0.7i$ and $g_P=0.63$ as these two points lie well inside the allowed regions of Figure~\ref{f:known}. With solutions in this region of parameter space we find $\tilde X,~\tilde Y,~\tilde Z$  are $ {\cal O}(10)$ to ${\cal O}(1000)$. It is important to emphasise, however, that these are only illustrations and that  there are infinitely many solutions. Looking at the four models then,

\begin{itemize}

\item Model I. We present numerical results for this case in Figure~\ref{f:mapto31}. On the left panel we illustrate the region where solutions exist in the $\tan\beta-m_{H^\pm}$  plane. 
We see that a lower value of $\tan\beta$ and/or $m_{H^\pm}$ is needed to obtain solutions with smaller values of 
$|\tilde\chi^{u,d,l}|$. The region shown is dominated by low values of $\tan\beta$ which are compatible with constraints from LHC and LEP on the {\it flavour conserving} version of this model as seen in Figure~4 of Ref.~\cite{Arbey:2017gmh}. Figure~9 of the same reference indicates that  values of $\tan\beta\lesssim 2$ are ruled out by $B$ decay constraints. These constraints, however, can be significantly modified by flavour changing parameters such as $\tilde\chi^d_{bs}$. We are not aware of any global fit to the full set of parameters in the general 2HDM. 

\begin{figure}[!h]
\includegraphics[width=4.5cm]{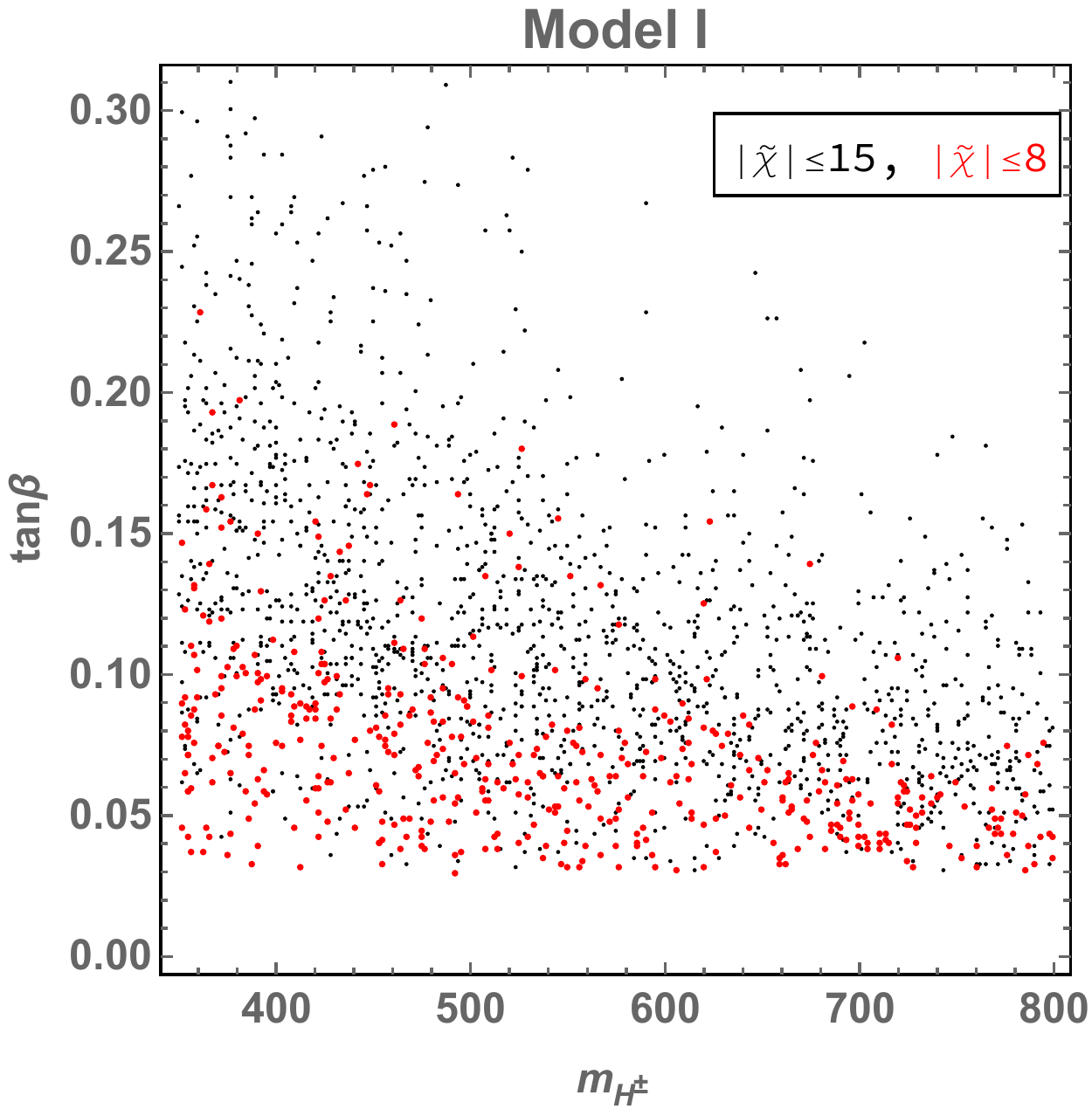}\includegraphics[width=4.5cm]{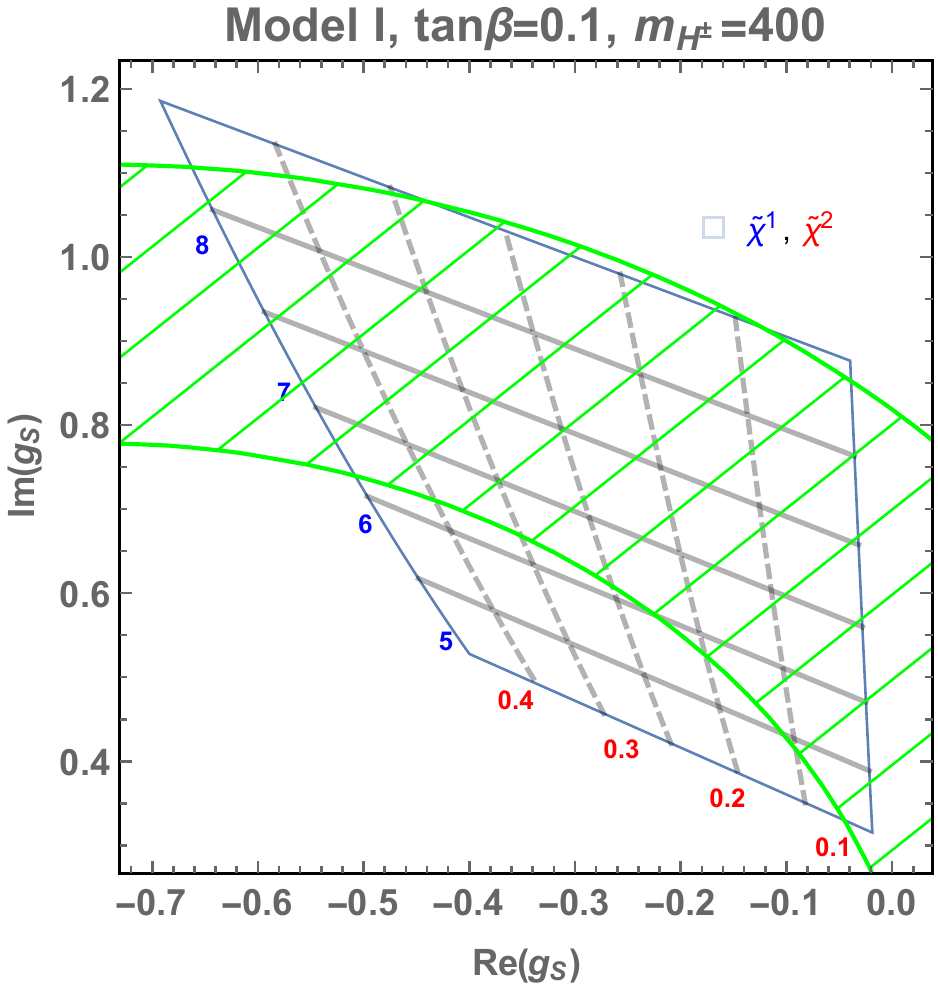}
\includegraphics[width=4.5cm]{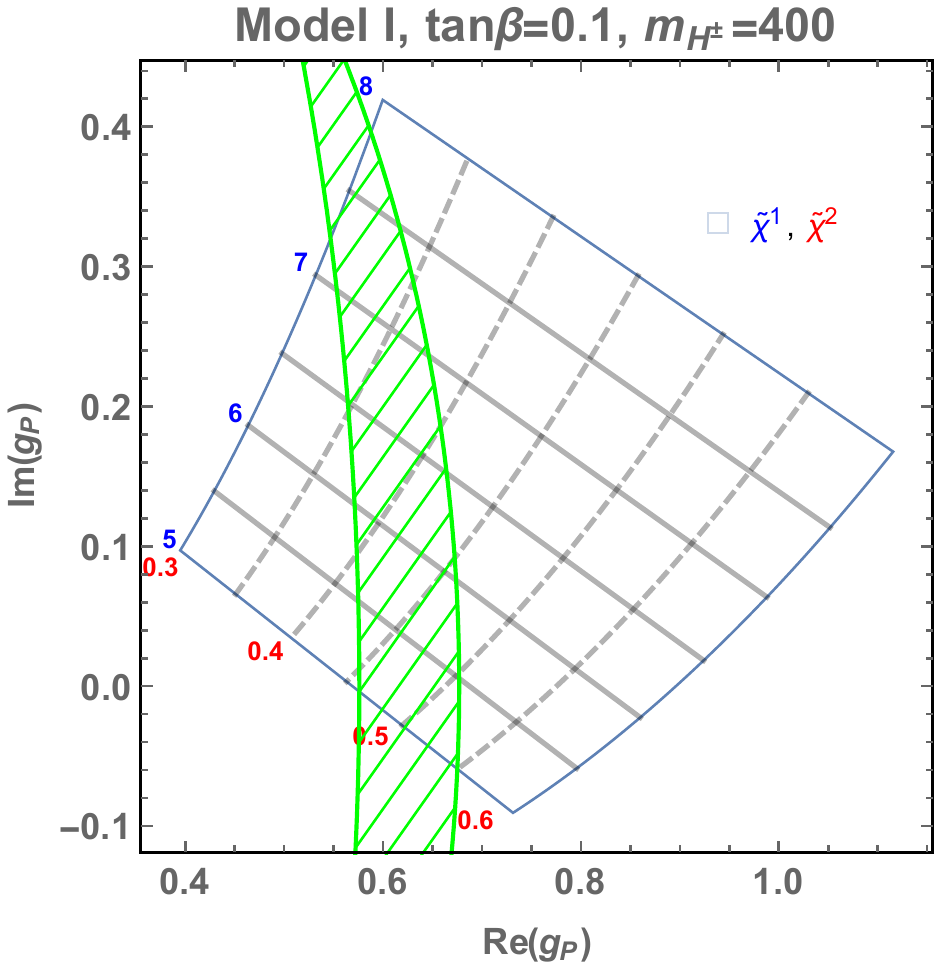}
\caption{Left panel: region where ${\cal R}(D^{(*)})$ agrees with experiment at $2\sigma$ and ${\mathcal{B}}(B_{c}\to\tau\nu)\leq30\%$ in the $\tan\beta-m_{H^\pm}$  plane. 
Centre and right panels: we map solutions with $\tilde\chi^d_{sb}=\tilde\chi^l_{\tau\tau}=\tilde\chi^1e^{0.8i}$, 
$\tilde\chi^u_{ct}=\tilde\chi^2e^{2i}$, $\tan\beta=0.1$ and $m_{H^\pm}=400$~GeV onto the allowed regions of Figure~\ref{f:known} shown in green.  }
\label{f:mapto31} 
\end{figure}

\item Model II. We present numerical results for this case in Figure~\ref{f:mapto32}. On the left panel we illustrate the region where solutions exist in the $\tan\beta-m_{H^\pm}$  plane. 
We see that in this case a higher value of $\tan\beta$ and/or a lower value of $m_{H^\pm}$ is needed to obtain solutions with smaller values of 
$|\tilde\chi^{u,d,l}|$. The $\tan\beta-m_{H^\pm}$ region of solutions in this case is consistent with the constraints on the corresponding  {\it flavour conserving} version of this model in Ref.~\cite{Arbey:2017gmh}. The centre and right panels illustrate that solutions consistent with the Cheng-Sher ansatz exist in this case.

\begin{figure}[!h]
\includegraphics[width=4.5cm]{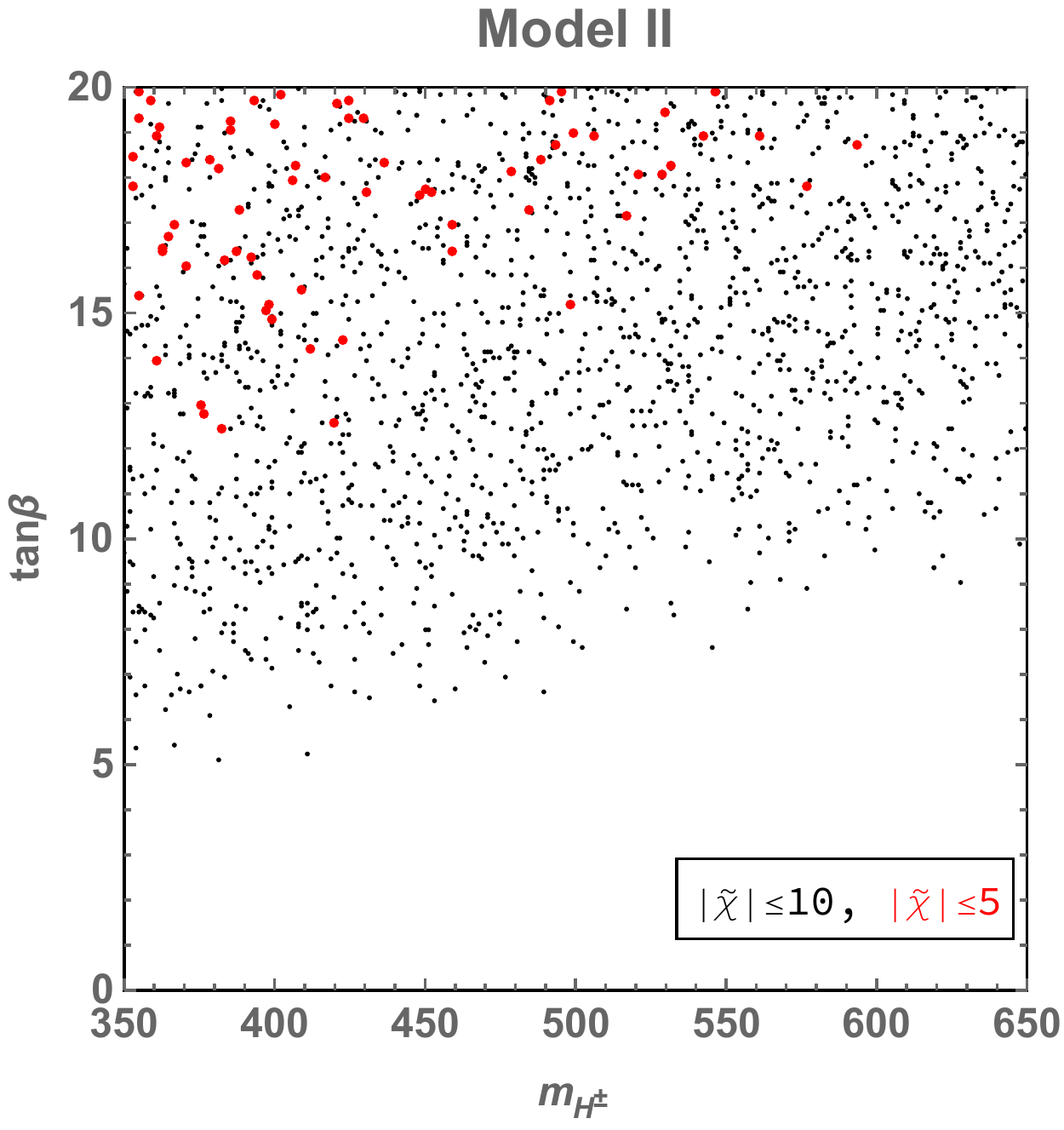}\includegraphics[width=4.5cm]{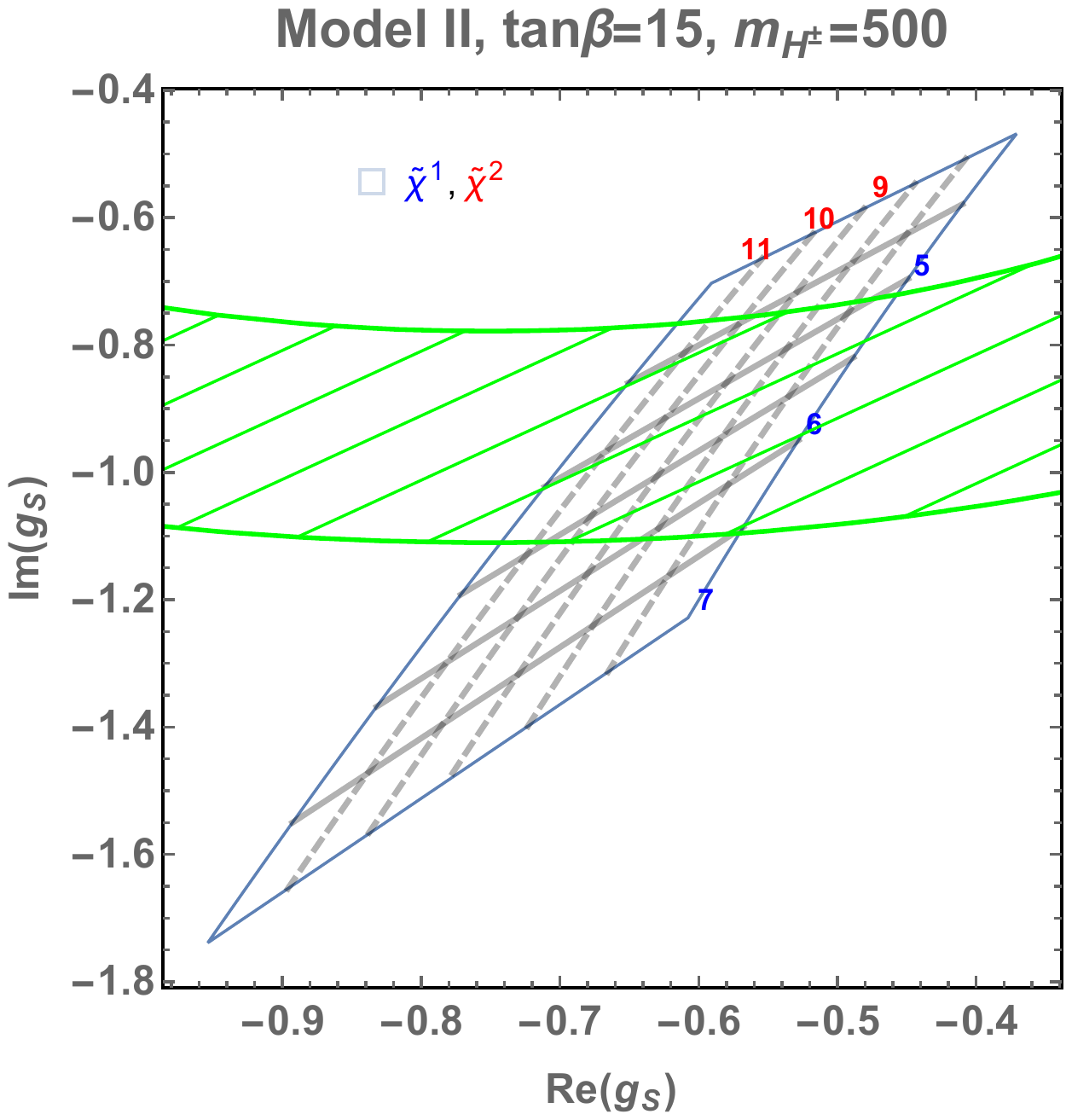}
\includegraphics[width=4.5cm]{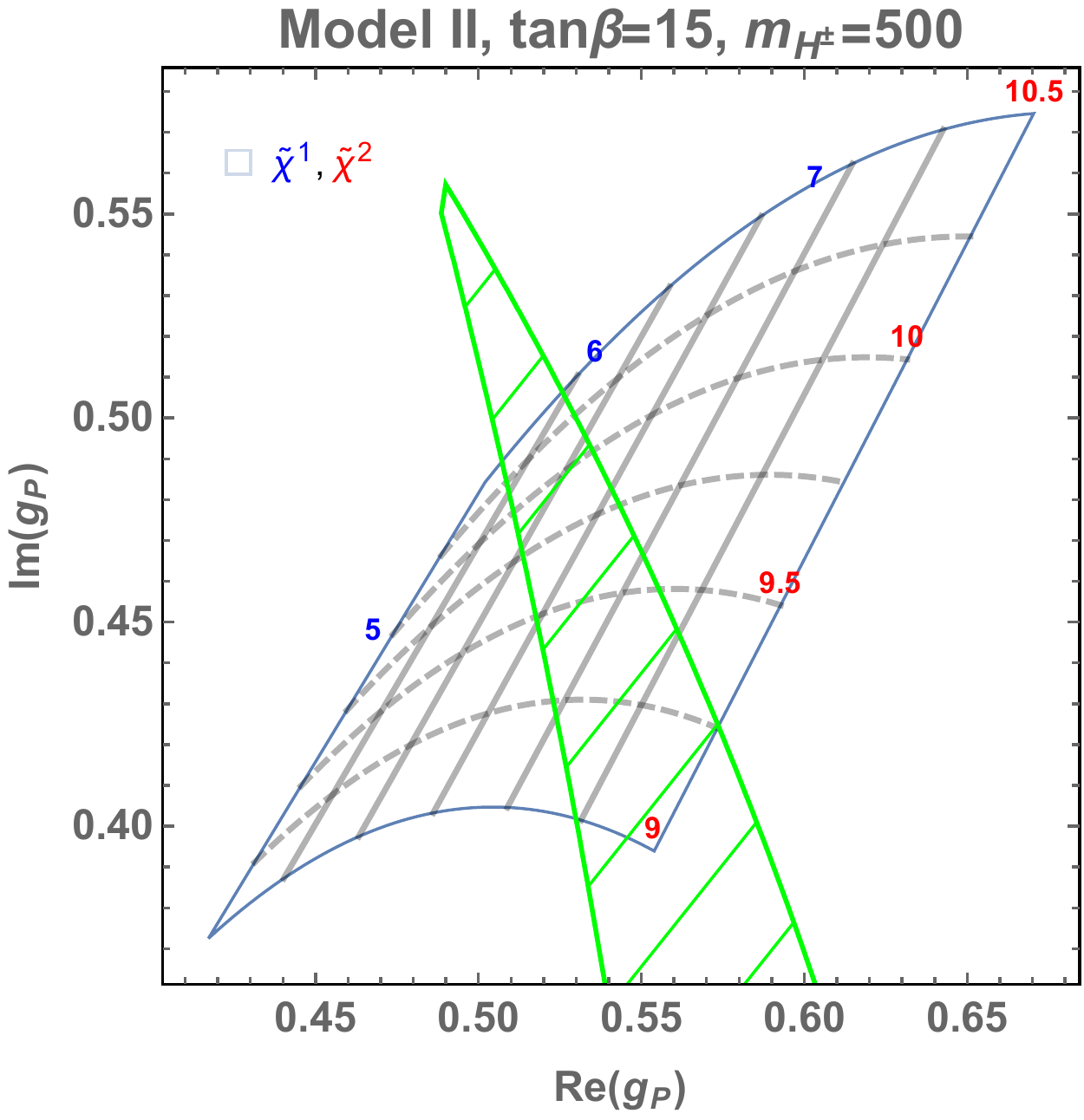}
\caption{Left panel: region where ${\cal R}(D^{(*)})$ agrees with experiment at $2\sigma$ and ${\mathcal{B}}(B_{c}\to\tau\nu)\leq30\%$ in the $\tan\beta-m_{H^\pm}$  plane. 
Centre and right panels: we map solutions with $\tilde\chi^d_{sb}=\tilde\chi^l_{\tau\tau}=\tilde\chi^1e^{-0.8i}$, 
$\tilde\chi^u_{ct}=\tilde\chi^2e^{-1.2i}$, $\tan\beta=15$ and $m_{H^\pm}=500$~GeV onto the allowed regions of Figure~\ref{f:known} shown in green.  }
\label{f:mapto32} 
\end{figure}

\item Model X. We present numerical results for this case in Figure~\ref{f:mapto3x}. On the left panel we illustrate the region where solutions exist in the $\tan\beta-m_{H^\pm}$  plane. 
We see that in this case a higher value of $\tan\beta$ and/or a lower value of $m_{H^\pm}$ is needed to obtain solutions with smaller values of 
$|\tilde\chi^{u,d,l}|$. This scenario is similar to Model~II in that the $\tan\beta-m_{H^\pm}$ region of solutions is consistent with the constraints on its corresponding  {\it flavour conserving} version as per Ref.~\cite{Arbey:2017gmh} (called type IV in that reference). The region illustrated on the centre and right panels needs $|\chi^{d,l}|$ values larger than what the Cheng-Sher ansatz would suggest are natural. However, the left panel indicates that there are other solutions which are also consistent with this ansatz.

\begin{figure}[!h]
\includegraphics[width=4.5cm]{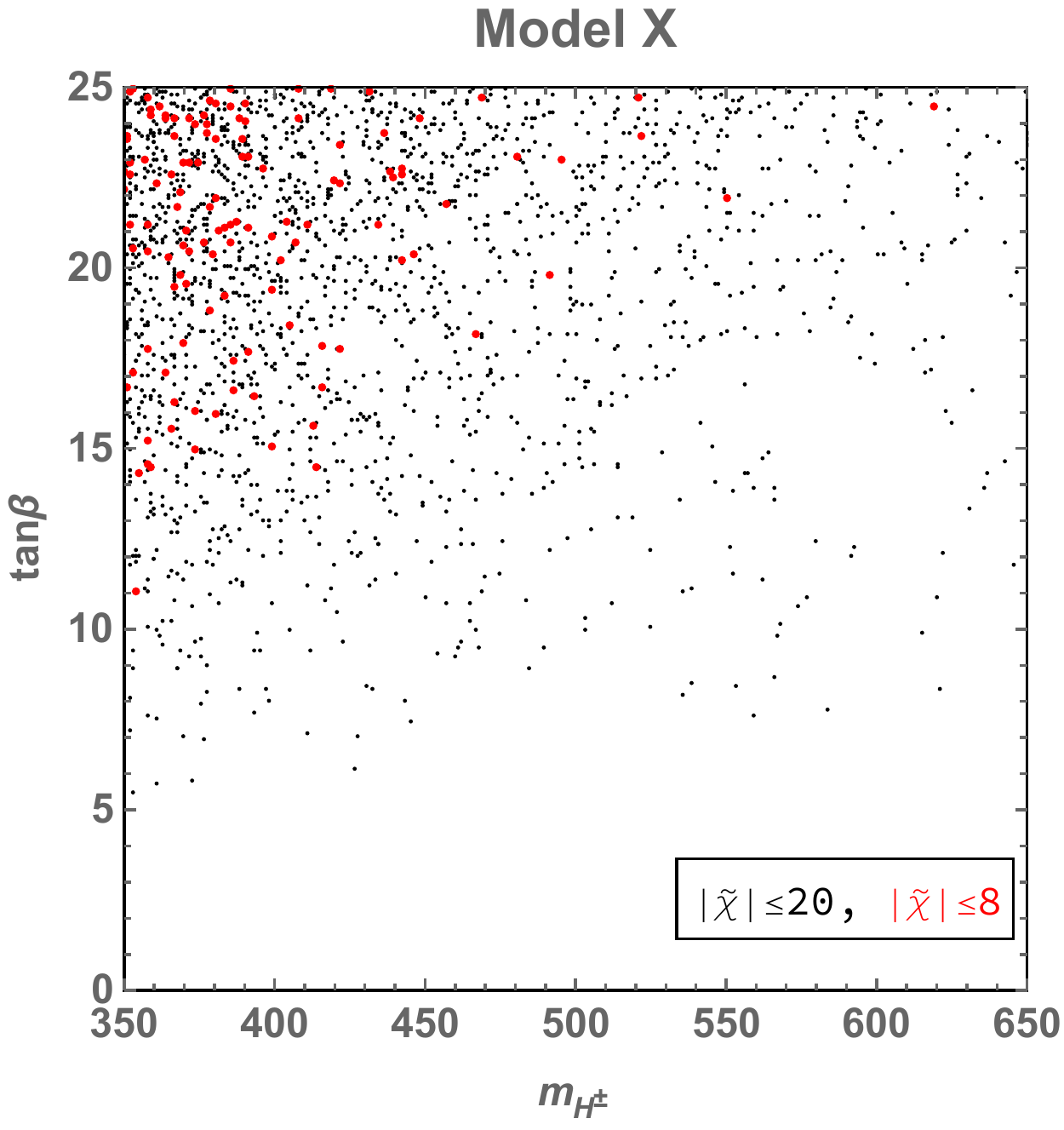}\includegraphics[width=4.5cm]{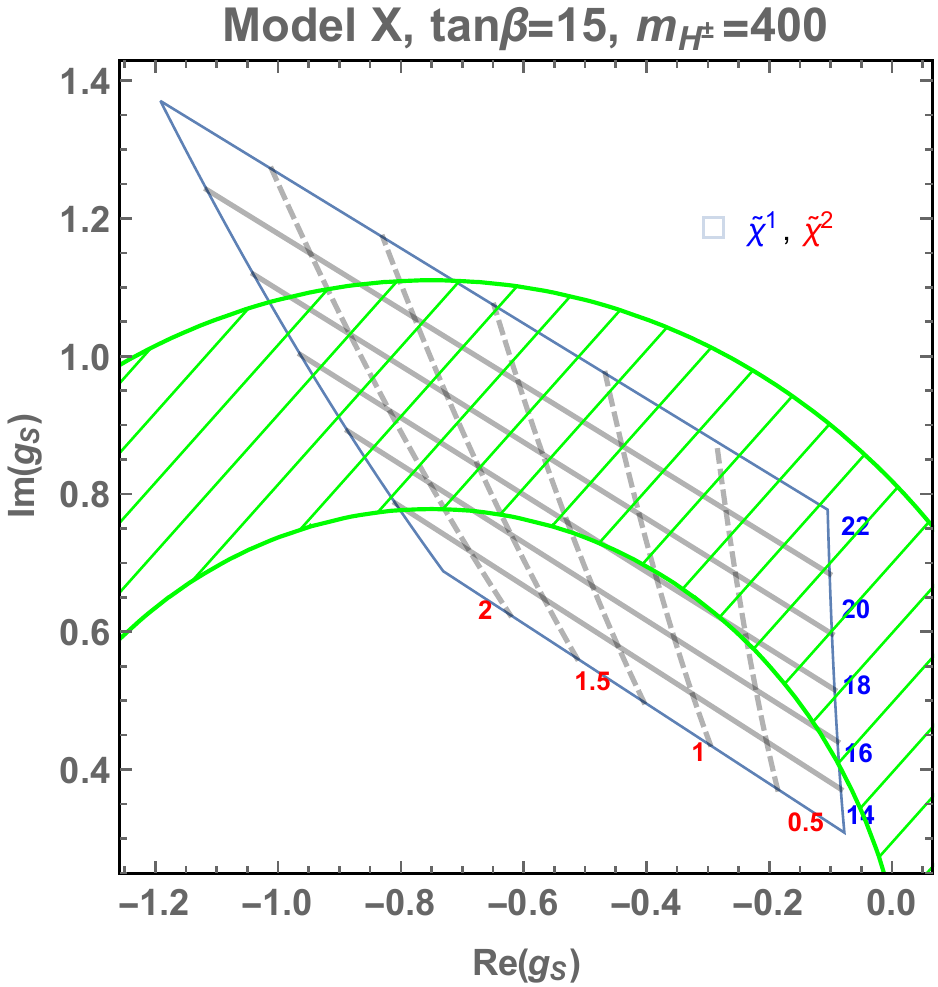}
\includegraphics[width=4.5cm]{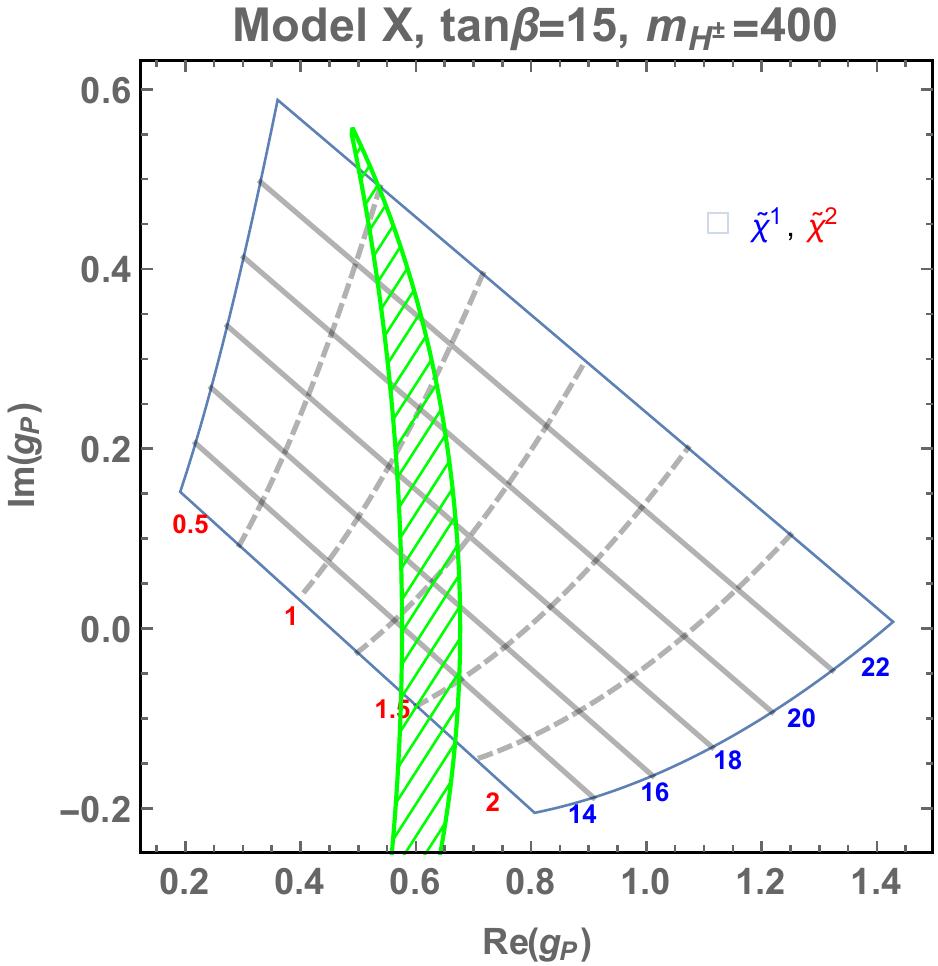}
\caption{Left panel: region where ${\cal R}(D^{(*)})$ agrees with experiment at $2\sigma$ and ${\mathcal{B}}(B_{c}\to\tau\nu)\leq30\%$ in the $\tan\beta-m_{H^\pm}$  plane. 
Centre and right panel: we map solutions with $\tilde\chi^d_{sb}=\tilde\chi^l_{\tau\tau}=\tilde\chi^1e^{-2.4i}$, 
$\tilde\chi^u_{ct}=\tilde\chi^2e^{-1.2i}$, $\tan\beta=15$ and $m_{H^\pm}=400$~GeV onto the allowed regions of Figure~\ref{f:known} shown in green.  }
\label{f:mapto3x} 
\end{figure}

\item Model Y. Finally, we present numerical results for this case in Figure~\ref{f:mapto3y}. On the left panel we illustrate the region where solutions exist in the $\tan\beta-m_{H^\pm}$  plane. 
We see that in this case a higher value of $\tan\beta$ and/or a lower value of $m_{H^\pm}$ is needed to obtain solutions with smaller values of 
$|\tilde\chi^{u,d,l}|$. The $\tan\beta-m_{H^\pm}$ region of solutions is once again consistent with its corresponding {\it flavour conserving} version~\cite{Arbey:2017gmh} (called type III in that reference), although the overlap region mostly lies in the upper range of both $\tan\beta$ and  $m_{H^\pm}$ shown in the left panel. This panel also suggests that in this case, the $\tilde\chi$ parameters are required to be larger than expected in the Cheng-Sher ansatz.

\begin{figure}[!h]
\includegraphics[width=4.5cm]{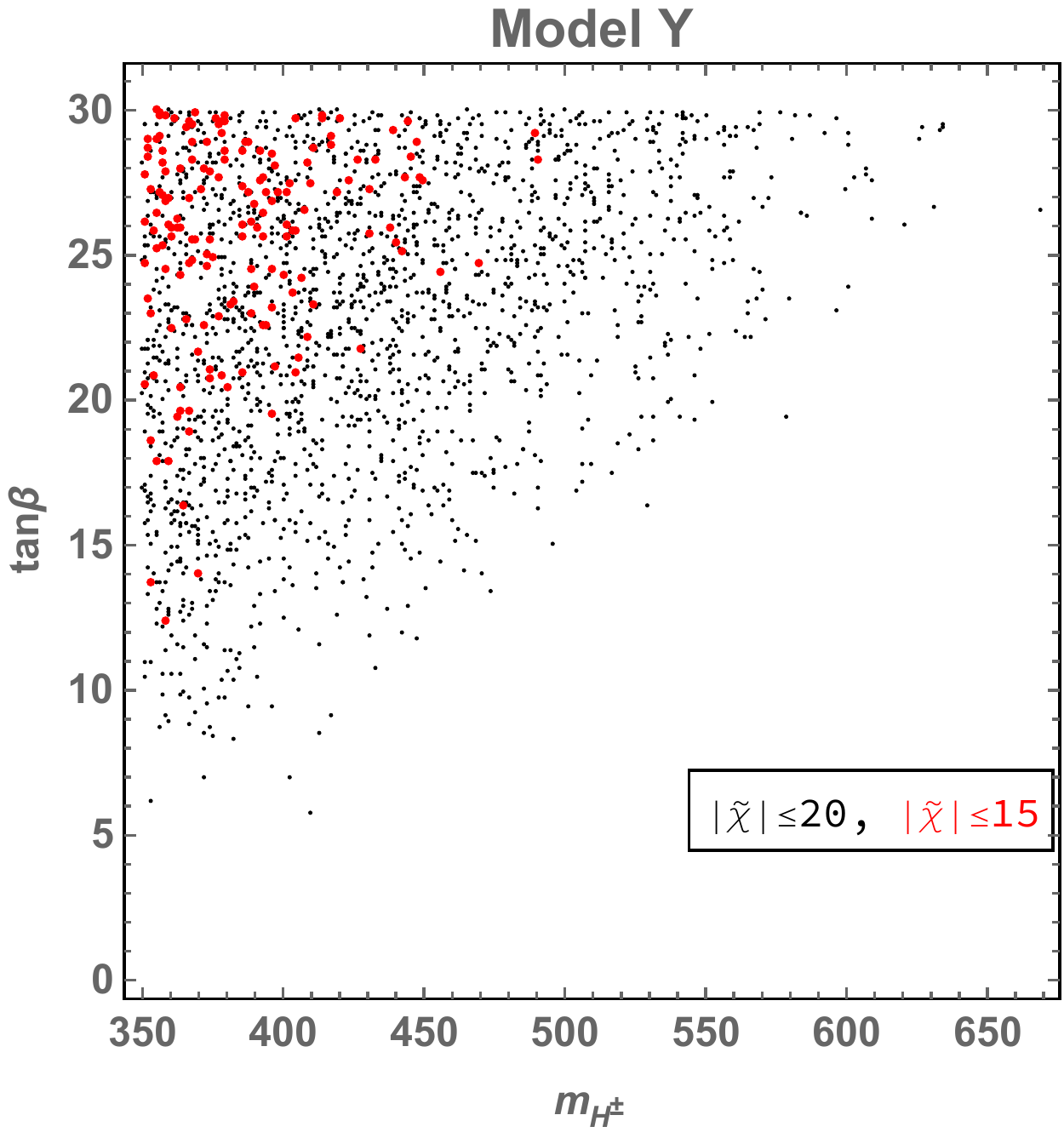}\includegraphics[width=4.5cm]{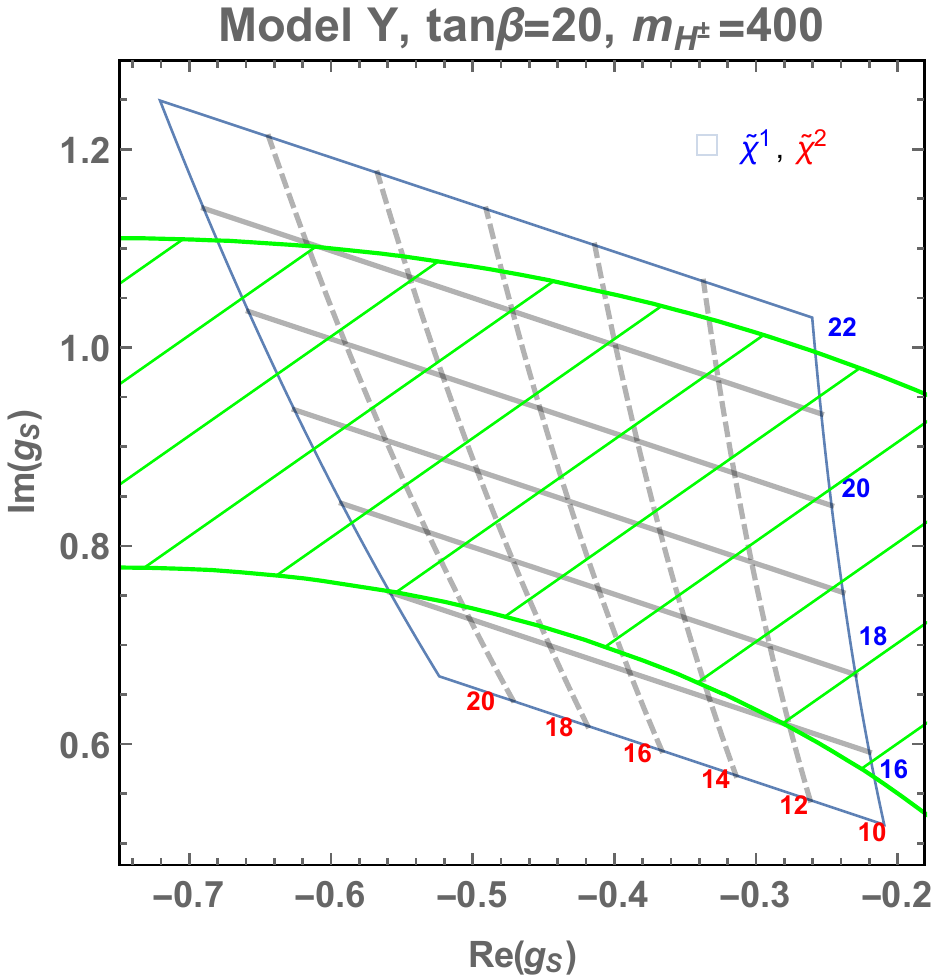}
\includegraphics[width=4.5cm]{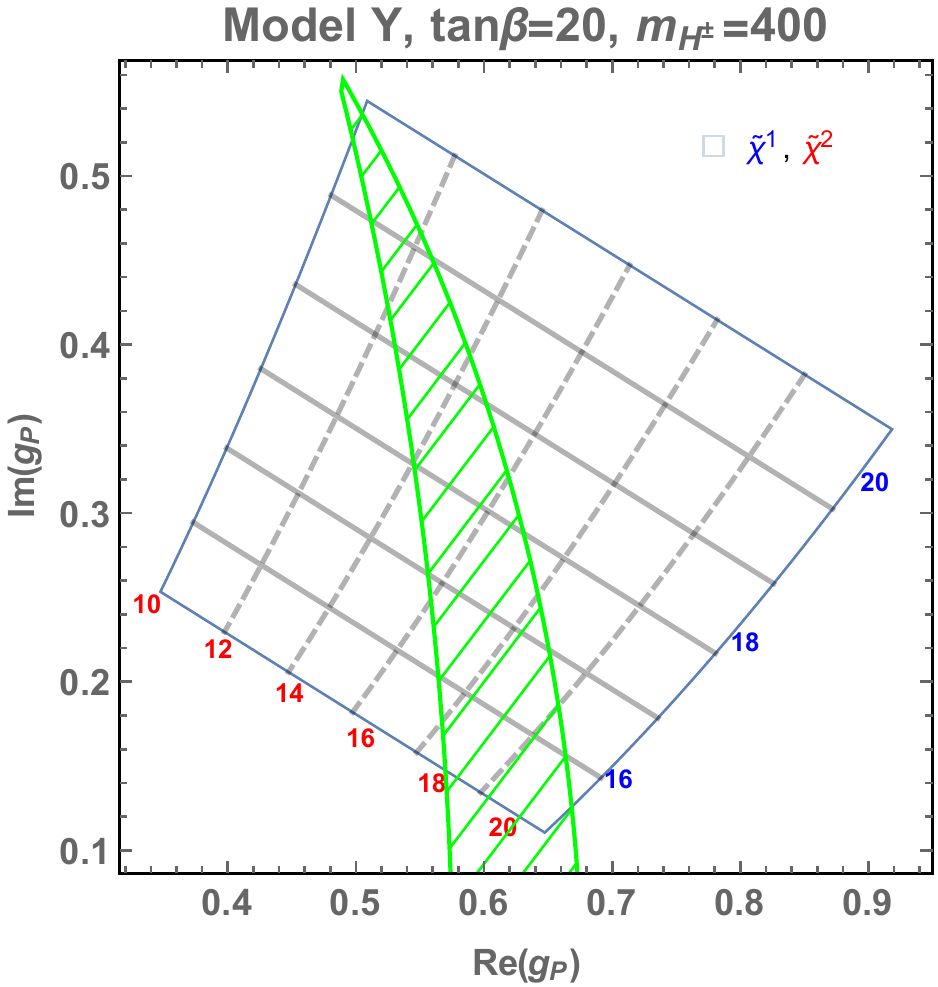}
\caption{Left panel: region where ${\cal R}(D^{(*)})$ agrees with experiment at $2\sigma$ and ${\mathcal{B}}(B_{c}\to\tau\nu)\leq30\%$ in the $\tan\beta-m_{H^\pm}$  plane. 
Centre (right) panel: we map solutions with $\tilde\chi^d_{sb}=\tilde\chi^l_{\tau\tau}=\tilde\chi^1e^{0.7i}$, 
$\tilde\chi^u_{ct}=\tilde\chi^2e^{2i}$, $\tan\beta=20$ and $m_{H^\pm}=400$~GeV onto the allowed regions of Figure~\ref{f:known} shown in green.   }
\label{f:mapto3y} 
\end{figure}
\end{itemize}

Additional considerations that may restrict the parameters in the
general model arise form Yukawa couplings to the neutral (SM-like)
Higgs defined as 
\begin{equation}
g_{hf_{i}f_{j}}=\frac{g}{2M_{W}}m_{f_{i}}h_{ij}^{f}.
\end{equation}
Once we introduce non-zero couplings $\tilde{\chi}_{ct}^{u},\ \tilde{\chi}_{ct}^{u},\ \tilde{\chi}_{\tau\tau}^{l}$
as in Eq.~\ref{cimplecou}, they also appear in $g_{h\tau\tau},\ g_{hct}$
and $g_{hsb}$, and are given by 
\begin{eqnarray}
h_{\tau\tau}^{l} & = & \begin{cases}
\frac{\cos\alpha}{\sin\beta}+\frac{\cos(\beta-\alpha)}{\sin\beta}\frac{\tilde{\chi}_{\tau\tau}^{l}}{\sqrt{2}} & \mbox{for~models I, Y }\\
-\frac{\sin\alpha}{\cos\beta}+\frac{\cos(\beta-\alpha)}{\cos\beta}\frac{\tilde{\chi}_{\tau\tau}^{l}}{\sqrt{2}} & \mbox{for~models II, X}
\end{cases}\nonumber \\
h_{ct}^{u} & = & -\frac{\cos(\beta-\alpha)}{\sin\beta}\frac{\tilde{\chi}_{ct}^{u}}{\sqrt{2}}\sqrt{\frac{m_{t}}{m_{c}}}\ \mbox{for~all~models}\nonumber \\
h_{sb}^{d} & = & \begin{cases}
\frac{\cos(\beta-\alpha)}{\sin\beta}\frac{\tilde{\chi}_{sb}^{d}}{\sqrt{2}}\sqrt{\frac{m_{b}}{m_{s}}} & \mbox{for~models I, X }\\
\frac{\cos(\beta-\alpha)}{\cos\beta}\frac{\tilde{\chi}_{sb}^{d}}{\sqrt{2}}\sqrt{\frac{m_{b}}{m_{s}}} & \mbox{for~models II, Y}
\end{cases}\label{fvc}
\end{eqnarray}
These expressions simplify in the alignment limit, defined as $\cos(\beta-\alpha)\to0$,
in which case the couplings of $h$ tend to the SM Higgs couplings.
To linear order in $\cos(\beta-\alpha)$ we obtain 
\begin{eqnarray}
|h_{\tau\tau}^{l}|^{2}-1 & \approx & \begin{cases}
-2\cos(\beta-\alpha)\left(\cot\beta+\frac{\tilde{\chi}_{\tau\tau}^{l}}{\sqrt{2}\sin\beta}\right) & \mbox{for~models I, Y }\\
2\cos(\beta-\alpha)\left(\tan\beta-\frac{\tilde{\chi}_{\tau\tau}^{l}}{\sqrt{2}\cos\beta}\right) & \mbox{for~models II, X}
\end{cases}
\end{eqnarray}

The first constraint arises from the process $h\rightarrow\tau^{+}\tau^{-}$,
for which the measured signal strength is \cite{Khachatryan:2016vau}
\begin{eqnarray}
\mu_{\tau}\equiv\frac{{\cal B}(h\rightarrow\tau^{+}\tau^{-})}{{\cal B}(h\rightarrow\tau^{+}\tau^{-})_{SM}}=1.11_{-0.22}^{+0.24},
\end{eqnarray}
leads to 
\begin{eqnarray}
-0.32\lesssim|h_{\tau\tau}^{l}|^{2}-1\lesssim0.58\label{chittcon}
\end{eqnarray}
at the 95\% confidence level.

In addition, if the flavour changing couplings get too large they
will conflict with the non-observation of $t\to ch$ and with indirect
limits on $h\to bs$. For $\mathcal{B}(t\to hc)<0.22\%$ at 95\% c.l.
\cite{Tanabashi:2018oca} one finds 
\begin{equation}
\left|\frac{\cos(\beta-\alpha)}{\sin\beta}\tilde{\chi}_{ct}^{u}\right|\lesssim1.4.
\end{equation}
The process $h\to bs$ has not been constrained yet, but it has been
argued in the literature that a branching ratio as large as $\mathcal{B}(h\to bs)\sim36\%$
can remain consistent with other flavour results in these types of
models \cite{Crivellin:2017upt}. Adopting this number and with the
95\% c. l. $\Gamma_{H}<0.013$~GeV \cite{Tanabashi:2018oca} we find, 
\begin{eqnarray}
|\frac{\cos(\beta-\alpha)}{\sin\beta}\tilde{\chi}_{sb}^{d}|\lesssim17\ \ \mbox{for~models I, X},\nonumber \\
|\frac{\cos(\beta-\alpha)}{\cos\beta}\tilde{\chi}_{sb}^{d}|\lesssim17\ \ \mbox{for~models II, Y}.\label{chibscon}
\end{eqnarray}
The constraints in Eqs.~\ref{chittcon}-\ref{chibscon} depend on
$\cos(\beta-\alpha)$ and disappear in the alignment limit. Ref.~\cite{Belusca-Maito:2016dqe}
presents upper bounds on $\cos(\beta-\alpha)$ of ${\cal O}(0.1)$
that depend on $\tan\beta$ for the four types of \textbf{flavor conserving}
models, so they do not automatically extend to our case.

\section{Summary Conclusions}

We have revisited the 2HDM-III as a possible explanation for the $\mathcal{R}(D^{(*)})$
anomalies. We first summarised the constraints known in the literature in terms of generic (pseudo)-scalar 
couplings and discussed the possible conflict between $\mathcal{R}(D^{*})$ and ${\mathcal{B}}(B_{c}\to\tau\nu)\leq30\%$.
We found that the parameter space that can explain these two
anomalies at the two-sigma level is limited to the region ${\mathcal{B}}(B_{c}\to\tau\nu)>23\%$. The bound ${\mathcal{B}}(B_{c}\to\tau\nu)>10\%$ 
advocated in Ref.\cite{Akeroyd:2017mhr} in turn restricts the possible explanation of $\mathcal{R}(D^{(*)})$ to the $3\sigma$ level within these models.

Armed with these constraints we predicted the ranges of other observables in  $b\to c \tau\nu$ reactions, including 
 $\mathcal{R}(J/\psi)$ and  $\mathcal{R}(\Lambda_{c})$. We find that the large central value in the current measurement of  $\mathcal{R}(J/\psi)$ 
 is consistent with this model at about the $2\sigma$ level with the currently large experimental error, but that a more precise measurement of 
 this quantity could place it in conflict with $\mathcal{R}(D^{(*)})$.
 
We found that the distributions $d\Gamma/dq^{2}$ in $B\to D\tau\nu$, $B\to D^{(*)}\tau\nu$ or $B_{c}\to J/\psi\tau\nu$ cannot distinguish between  the SM or models 
with new (pseudo)-scalar couplings. 

We presented predictions for the tau-lepton polarisation in $B\to D^{(*)}\tau\nu$
in the presently allowed region of parameter space. In particular we find that phases in the Yukawa couplings  can produce substantial T-odd normal polarisations.
 
We considered four versions of the 2HDM-III which are constructed by extending 
the four flavour conserving 2HDM with the addition of flavour
changing couplings that we have limited in size with the Sher-Cheng ansatz. We mapped the allowed regions in $g_P-g_S$ into the parameter space of
these four models. We found that the allowed $(m_{H^{\pm}},\tan\beta)$
ranges also satisfy the LHC and LEP constraints found in the
literature for the {\it flavour conserving} versions of these models. We also found that the allowed regions of parameter space are
not further constrained by $h\to\tau\tau$, $t\to hc$, $h\to bs$.

\appendix
\section{Helicity Amplitudes }

The invariant form factors describing the hadronic transitions $\bar{B}\to D$
and $\overline{B}\to D^{*}$ are defined as usual 
\begin{eqnarray}
 &  & \langle D(p_{2})|\bar{c}\gamma^{\mu}b|\bar{B}(p_{1})\rangle=F_{+}(q^{2})P^{\mu}+F_{-}(q^{2})q^{\mu},\nonumber \\
 &  & \langle D(p_{2})|\bar{c}b|\bar{B}(p_{1})\rangle=(m1+m2)F^{S}(q^{2}),\nonumber \\
 &  & \langle D^{*}(p_{2})|\bar{c}\gamma^{\mu}(1\mp\gamma^{5})b|\bar{B}(p_{1})\rangle=\frac{\epsilon_{2\alpha}^{\dagger}}{m_{1}+m_{2}}\Big[\mp g^{\mu\alpha}PqA_{0}(q^{2})\nonumber \\
 &  & \pm P^{\mu}P^{\alpha}A_{+}(q^{2})\pm q^{\mu}P^{\alpha}A_{-}(q^{2})+i\varepsilon^{\mu\alpha}PqV(q^{2})\Big],\nonumber \\
 &  & \langle D^{*}(p_{2})|\bar{c}\gamma^{5}b|\bar{B}(p_{1})\rangle=\epsilon_{2\alpha}^{\dagger}P^{\alpha}G^{S}(q^{2}),\label{defff}
\end{eqnarray}
where $P=p_{1}+p_{2}$, $q=p_{1}-p_{2}$, and $\epsilon_{2}$ is the
polarization vector of the $D^{*}$ meson which satisfies $\epsilon_{2}^{\dagger}\cdot p_{2}=0$.
The particles are on their mass shells: $p_{1}^{2}=m_{B}^{2}$ and
$p_{2}^{2}=m_{D^{(\ast)}}^{2}$. All the expressions are written in
 terms of helicity form factors, which are related to those in Eq.~\ref{defff}
for the $\overline{B}\to D$ transition by \citep{Ivanov:2017mrj}
 \begin{eqnarray}
H_{t} & = & \frac{PqF_{+}+q^{2}F_{-}}{\sqrt{q^{2}}},\,\\
H_{0} & = & \frac{2m_{B}|\mathbf{p_{2}|}F_{+}}{\sqrt{q^{2}}},\\
H_{P}^{S} & = & (m_{B}+m_{D})F^{S},
\end{eqnarray}
where $|{\bf p_{2}}|=\lambda^{1/2}(m_{B}^{2},m_{D^{(*)}}^{2},q^{2})/2m_{B}$
is the momentum of the daughter meson with $\lambda=(m_{B}^{2}-m_{D^{(*)}}^{2}-q^{2})^{2}-4m_{D^{(*)}}^{2}q^{2}.$
For the $\overline{B}\to D^{*}$ transition \citep{Ivanov:2017mrj}
 \begin{eqnarray}
H_{t} & = & \frac{m_{B}|{\bf p_{2}}|\left(Pq(-A_{0}+A_{+})+q^{2}A_{-}\right)}{m_{D^{*}}\sqrt{q^{2}}(m_{B}+m_{D^{*}})},\\
H_{\pm} & = & \frac{-PqA_{0}\pm2m_{1}|{\bf p_{2}}|V}{m_{B}+m_{D^{*}}},\\
H_{V}^{S} & = & \frac{m_{B}}{m_{D^{*}}}|{\bf p_{2}}|G^{S},\\
H_{0} & = & \frac{-Pq(m_{B}^{2}-m_{D^{*}}^{2}-q^{2})A_{0}+4m_{B}^{2}|{\bf p_{2}}|^{2}A_{+}}{2m_{2}\sqrt{q^{2}}(m_{B}+m_{D^{*}})}.
\end{eqnarray}

For our numerical estimates we use the helicity amplitudes calculated
in the covariant confined quark model (CCQM) with the double-pole
parameterisation of Refs.~\cite{Ivanov:2017mrj,Ivanov:2016qtw}:
 \begin{eqnarray}
j(q^{2}) & = & \frac{j(0)}{1-a_{j}s+b_{j}s^{2}},\quad s=\frac{q^{2}}{m_{B}^{2}},\nonumber \\
 &  & j=F_{\pm},\,A_{0,\pm},\,F^{S},\,G^{S},\,V.
\end{eqnarray}
Similarly, for the $\Lambda_{b}$ decay we need the vector and axial
current form factors \cite{Li:2016pdv,Gutsche:2015mxa} 
 \begin{equation}
H_{\lambda_{2},\lambda_{W}}=H_{\lambda_{2},\lambda_{W}}^{V}-H_{\lambda_{2},\lambda_{W}}^{A},
\end{equation}
\begin{equation}
H_{\lambda_{2},\lambda_{W}}^{V(A)}=\epsilon^{\dagger\mu}(\lambda_{W})\langle\Lambda_{c},\lambda_{2}|\bar{c}\gamma_{\mu}(\gamma_{\mu}\gamma_{5})b|\Lambda_{b},\lambda_{1}\rangle,\label{eq:V-A amps}
\end{equation}
which satisfy the parity relations,

\begin{equation}
H_{\lambda_{2},\lambda_{W}}^{V}=H_{-\lambda_{2},-\lambda_{W}}^{V},\quad H_{\lambda_{2},\lambda_{W}}^{A}=-H_{-\lambda_{2},-\lambda_{W}}^{A},
\end{equation}

\begin{equation}
H_{-\lambda_{2},-\lambda_{W}}=H_{\lambda_{2},\lambda_{W}}^{V}+H_{\lambda_{2},\lambda_{W}}^{A},
\end{equation}
where $\lambda_{2}$ and $\lambda_{W}$ denote the helicities of the
daughter baryon $\Lambda_{c}$ and the virtual $W$ boson respectively.
In the SM the helicity amplitudes $H_{\lambda_{2},\lambda_{W}}^{V(A)}$are
given by \cite{Gutsche:2015mxa}

\begin{align}
H_{+\frac{1}{2},t}^{V(A)} & =\sqrt{\frac{Q_{\pm}}{q^{2}}}\left(M_{\mp}F_{1}^{V(A)}\pm\frac{q^{2}}{M_{\Lambda_{b}}}F_{3}^{V(A)}\right),\\
H_{+\frac{1}{2},+}^{V(A)} & =\sqrt{2Q_{\mp}}\left(F_{1}^{V(A)}\pm\frac{M_{\pm}}{M_{\Lambda_{b}}}F_{2}^{V(A)}\right),\\
H_{+\frac{1}{2},0}^{V(A)} & =\sqrt{\frac{Q_{\mp}}{q^{2}}}\left(M_{\pm}F_{1}^{V(A)}\pm\frac{q^{2}}{M_{\Lambda_{b}}}F_{2}^{V(A)}\right).
\end{align}

\noindent with $M_{\pm}=M_{\Lambda_{b}}\pm M_{\Lambda_{c}}$ and $Q_{\pm}=M_{\pm}^{2}-q^{2}$.
The helicity amplitudes for scalar and pseudo-scalar operators needed
for 2HDM are \cite{Shivashankara:2015cta} 
 \begin{align}
H_{\lambda_{2},0}^{SP}= & H_{\lambda_{2},0}^{S}-H_{\lambda_{2},0}^{P},\label{eq:S-P amps}\\[0.2cm]
H_{\pm\frac{1}{2},0}^{SP}= & \frac{\sqrt{Q_{+}}}{m_{b}-m_{c}}\left(M_{-}F_{1}^{V}+\frac{q^{2}}{M_{\Lambda_{b}}}F_{3}^{V}\right)\nonumber \\
 & \pm\frac{\sqrt{Q_{-}}}{m_{b}+m_{c}}\left(M_{+}F_{1}^{A}-\frac{q^{2}}{M_{\Lambda_{b}}}F_{3}^{A}\right).
\end{align}

In this way, the partial decay width of the $\Lambda_{b}\to\Lambda_{c}l\bar{\nu}$
process is given by \citep{Shivashankara:2015cta}

\begin{equation}
\frac{{\rm d}\Gamma}{{\rm d}q^{2}}=\frac{G_{F}^{2}|V_{cb}|^{2}q^{2}|\vec{\mathbf{p}}_{2}|}{192\pi{}^{3}M_{\Lambda_{b}}^{2}}\left(1-2\delta_{l}\right)^{2}\left[\left(1+\delta_{l}\right)\sum_{ij}H_{ij}^{2}+\frac{3}{2}B_{3}^{NP}+\frac{3m_{l}}{\sqrt{q^{2}}}B_{4}^{Int}\right]\label{eq:lambda decay width}
\end{equation}
where $\delta_{l}=m_{l}^{2}/2q^{2}$, $ij=(-\frac{1}{2},-),\,(-\frac{1}{2},0),\,(\frac{1}{2},0),\,(\frac{1}{2},+)$
and
 \begin{eqnarray}
B_{3}^{NP} & = & |H_{1/2,0}^{SP}|^{2}+|H_{-1/2,0}^{SP}|^{2},\nonumber \\
B_{4}^{Int} & = & \mathrm{Re}(H_{1/2,t}\ (H_{1/2,0}^{SP})^{*}+H_{-1/2,t}\ (H_{-1/2,0}^{SP})^{*}).
\end{eqnarray}

\section{Polarisations }

Following Ref.~\cite{Ivanov:2017mrj}, the ratios ${\cal R}(D^{(*)})$
 are given by 
\begin{equation}
\mathcal{R}(D^{(*)})=\frac{\left(\frac{q^{2}-m_{\tau}^{2}}{q^{2}-m_{\mu}^{2}}\right){\cal H}_{tot}(D^{(*)})}{\sum_{n}|H_{n}|^{2}+\delta_{\mu}\big(\sum_{n}|H_{n}|^{2}+3|H_{t}|^{2}\big)},
\end{equation}
with 
\begin{eqnarray}
{\cal H}_{tot}(D) & =\left[|H_{0}|^{2}+\delta_{\tau}(|H_{0}|^{2}+3|H_{t}|^{2}\right]+\frac{3}{2}|g_{S}|^{2}|H_{P}^{S}|^{2}+3\sqrt{2\delta_{\tau}}\textrm{Re[}g_{S}]H_{P}^{S}H_{t},\nonumber \\
{\cal H}_{tot}(D^{*}) & =\sum_{n}|H_{n}|^{2}\left(\delta_{\tau}+1\right)+3\delta_{\tau}|H_{t}|^{2}+\frac{3}{2}|g_{P}|^{2}|H_{V}^{S}|^{2}-3\sqrt{2\delta_{\tau}}\textrm{Re[}g_{P}]H_{V}^{S}H_{t},\label{Htot D*}
\end{eqnarray}
and $g_{S}\equiv\left(C_{L}^{cb}+C_{R}^{cb}\right)/C_{SM}^{cb}$ and
$g_{P}\equiv\left(C_{L}^{cb}-C_{R}^{cb}\right)/C_{SM}^{cb}$. In terms
of Eq.~\ref{Htot D*}, the longitudinal differential polarisation
will be, 
 \begin{eqnarray}
{\cal H}_{tot}(D)\frac{dP_{L}^{\tau}(D)}{dq^{2}} & = & \left[|H_{0}|^{2}\left(\delta_{\tau}-1\right)+3\delta_{\tau}|H_{t}|^{2}+\frac{3}{2}|g_{S}|^{2}|H_{P}^{S}|^{2}+3\sqrt{2\delta_{\tau}}\textrm{Re(}g_{S})H_{P}^{S}H_{t}\right],\label{eq:PLD*}\\
{\cal H}_{tot}(D^{*})\frac{dP_{L}^{\tau}(D^{*})}{dq^{2}} & = & \left[\sum_{n}|H_{n}|^{2}\left(\delta_{\tau}-1\right)+3\delta_{\tau}|H_{t}|^{2}+\frac{3}{2}|g_{P}|^{2}|H_{V}^{S}|^{2}-3\sqrt{2\delta_{\tau}}\textrm{Re(}g_{P})H_{V}^{S}H_{t}\right].\nonumber 
\end{eqnarray}
Similarly, the transverse polarisation is given by 
\begin{eqnarray}
\frac{dP_{T}^{\tau}(D)}{dq^{2}} & = & \frac{3\pi\sqrt{\delta_{\tau}}}{2\sqrt{2}{\cal H}_{tot}(D)}\left(H_{0}H_{t}+\frac{\textrm{Re(}g_{S})H_{P}^{S}H_{0}}{\sqrt{2\delta_{\tau}}}\right),\nonumber \\
\frac{dP_{T}^{\tau}(D^{*})}{dq^{2}} & = & \frac{3\pi\sqrt{\delta_{\tau}}}{4\sqrt{2}{\cal H}_{tot}(D^{*})}\left[\left(|H_{-}|^{2}-|H_{+}|^{2}\right)+2H_{0}H_{t}-\frac{2\textrm{Re}(g_{P})H_{V}^{S}H_{0}}{\sqrt{2\delta_{\tau}}}\right].\label{eq:PTD*}
\end{eqnarray}
Finally, in the presence of CP-violating phases in the NP Higgs exchange
amplitude, there is a normal differential polarisation that reads \footnote{Note that there is a typo in Eq.~41 of Ref.~\cite{Ivanov:2017mrj} where the denominator of $P_{N}^{(D)}(q^2)$ should have a 4 instead of a 2. We thank C. T. Tran for confirming this.}
\begin{eqnarray}
\frac{dP_{N}^{\tau}(D)}{dq^{2}} & = & \frac{-3\pi}{4{\cal H}_{tot}(D)}\textrm{Im(}g_{S})H_{P}^{S}H_{0},\nonumber \\
\frac{dP_{N}^{\tau}(D^{*})}{dq^{2}} & = & \frac{3\pi}{4{\cal H}_{tot}(D^{*})}\textrm{Im(}g_{P})H_{V}^{S}H_{0}.\label{eq:PND*}
\end{eqnarray}

To calculate the integrated, or $q^{2}$ averaged polarisations, one
has to include the $q^{2}$-dependent phase-space factor $C(q^{2})=|\mathbf{p_{2}}|(q^{2}-m_{\tau}^{2})^{2}/q^{2}$
\cite{Ivanov:2017mrj}, 
\begin{equation}
P_{i}^{\tau}(D^{(*)})=\frac{\intop_{m_{\tau}^{2}}^{q_{max}^{2}}dq^{2}C(q^{2})\frac{dP_{i}^{\tau}(D^{(*)})}{dq^{2}}{\cal H}_{tot}(D^{(*)})}{\intop_{m_{\tau}^{2}}^{q_{max}^{2}}dq^{2}C(q^{2}){\cal H}_{tot}(D^{(*)})}.\label{eq:averagePol}
\end{equation}

\section*{Acknowledgments}

This work was partially supported by El Patrimonio Aut\'onomo Fondo
Nacional de Financiamiento para la Ciencia, la Tecnolog\'ia y la Innovaci\'on
Francisco Jos\'{e} de Caldas, COLCIENCIAS, Colombia. G.V. thanks the Departamento
de F\'isica, Universidad Nacional de Colombia for their hospitality
and partial support when this work was started. C. F. Sierra wants
to acknowledge J. Cuadros for her valuable support and advise. We
thank Ken Kiers who pointed out an error in the original manuscript.

\newpage

\bibliography{biblio.bib}

\end{document}